\documentclass[
reprint,
superscriptaddress,
amsmath,
prx,
]{revtex4-1}

\usepackage{graphicx}
\usepackage{bm}

\begin{document}

\title{Intertwined Orders in Heavy-Fermion Superconductor CeCoIn$_5$}

\author{Duk Y. Kim}
\email[]{dykim@lanl.gov}
\affiliation{Los Alamos National Laboratory, Los Alamos, New Mexico 87545, USA}

\author{Shi-Zeng Lin}
\affiliation{Los Alamos National Laboratory, Los Alamos, New Mexico 87545, USA}

\author{Franziska Weickert}
\affiliation{Los Alamos National Laboratory, Los Alamos, New Mexico 87545, USA}

\author{Michel Kenzelmann}
\affiliation{Laboratory for Scientific Developments \& Novel Materials, Paul Scherrer Institute, Villigen CH-5232, Switzerland}

\author{Eric D. Bauer}
\affiliation{Los Alamos National Laboratory, Los Alamos, New Mexico 87545, USA}

\author{Filip Ronning}
\affiliation{Los Alamos National Laboratory, Los Alamos, New Mexico 87545, USA}

\author{J. D. Thompson}
\affiliation{Los Alamos National Laboratory, Los Alamos, New Mexico 87545, USA}

\author{Roman Movshovich}
\email[]{roman@lanl.gov}
\affiliation{Los Alamos National Laboratory, Los Alamos, New Mexico 87545, USA}

\begin{abstract}
The appearance of spin-density-wave (SDW) magnetic order in the low-temperature and high-field corner of the superconducting phase diagram of CeCoIn$_5$ is unique among unconventional superconductors. The nature of this magnetic $Q$ phase is a matter of current debate. Here, we present the thermal conductivity of CeCoIn$_5$ in a rotating magnetic field, which reveals the presence of an additional order inside the $Q$ phase that is intimately intertwined with the superconducting $d$-wave and SDW orders. A discontinuous change of the thermal conductivity within the $Q$ phase, when the magnetic field is rotated about antinodes of the superconducting $d$-wave order parameter, demands that the additional order must change abruptly together with the recently observed switching of the SDW. A combination of interactions, where spin-orbit coupling orients the SDW, which then selects the secondary $p$-wave pair-density-wave component (with an average amplitude of 20\% of the primary $d$-wave order parameter), accounts for the observed behavior.
\end{abstract}

\maketitle

\section{Introduction}

Magnetism is considered to be detrimental to conventional superconductivity, which is mediated by lattice vibrations---phonons \cite{BCS}. An external magnetic field, for example, destroys superconductivity via either orbital \cite{Hohenberg} or spin \cite{Clogston} (Pauli) limiting mechanisms. A growing number of cases, however, display the coexistence of magnetism and superconductivity and constitute a fascinating problem in condensed-matter physics \cite{Lake2002,Cai2013}. CeCoIn$_5$ presents a unique case among all unconventional superconductors wherein a novel magnetic state, the so-called $Q$ phase, develops at high fields and requires superconductivity for its very existence. This $Q$ phase was originally suggested \cite{Bianchi2002,Bianchi2003,Radovan2003} to be a realization of spatially inhomogeneous superconductivity, the Fulde-Ferrell-Larkin-Ovchinnikov (FFLO) state \cite{FF,LO}. Subsequent NMR \cite{Young2007} and neutron scattering measurements \cite{Kenzelmann2008,Kenzelmann2010} revealed the presence of a magnetic spin-density-wave (SDW) order in the $Q$ phase. A number of theories were proposed for its origin \cite{Agterberg,Aperis,Yanase2009,Yanase2011,Michal2011,Hatakeyama2011,Martiny}, many of them involving additional orders, distinct from the $d$-wave superconductivity \cite{Izawa2001,Zhou2013} and the SDW. The issue of intertwined orders (magnetic, multiple and inhomogeneous superconductivity, etc.) is increasingly common in correlated systems \cite{Fradkin2015,Hamidian2016}. The $Q$ phase is a model system for studying such intertwined orders, with a uniquely tunable single-domain structure due to the high purity of CeCoIn$_5$.

Experimentally, neutron-scattering measurements suggest the condensation of a superconducting spin resonance as a possible origin \cite{Michal2011,Stock2008,Raymond2015} of the $Q$ phase. Recent neutron-scattering measurements reveal that its SDW order is single domain, with the ordering wave vector $\mathbf{Q}_\mathrm{SDW}$ either $\mathbf{Q}_1 = (q, q, 0.5)$ or $\mathbf{Q}_2 = (q, -q, 0.5)$, with $q\approx0.44$ along the two nodal directions of the superconducting $d$-wave order parameter \cite{Gerber2014}. When the magnetic field is rotated within the crystallographic $ab$ plane about the [100] direction, $\mathbf{Q}_\mathrm{SDW}$ switches abruptly between $\mathbf{Q}_1$ and $\mathbf{Q}_2$, choosing the one that is more perpendicular to the magnetic field [Figs. 1(b) and 1(c)]. It was suggested that a secondary $p$-wave pair-density-wave (PDW) component drives the hypersensitivity of $\mathbf{Q}_\mathrm{SDW}$ to the direction of the magnetic field \cite{Gerber2014}. This mechanism, however, at present lacks theoretical support (see Appendix A). One recently proposed scenario explains the hypersensitivity as being due to the magnetic field lifting the degeneracy of the direction of $\mathbf{Q}_\mathrm{SDW}$ via spin-orbit coupling \cite{Mineev2015}, without requiring any additional order besides the existing superconducting $d$-wave and SDW orders. Yet another scenario introduces the spatially inhomogeneous Fulde-Ferrell-Larkin-Ovchinnikov (FFLO) state, which couples to the SDW state and lowers the energy of the $Q$ phase with $\mathbf{Q}_\mathrm{SDW}$ more perpendicular to $\bm{q}_{\mathrm{FFLO}}$ \cite{Hatakeyama2015}. The mechanism responsible for the switching of the direction of $\mathbf{Q}_\mathrm{SDW}$ is a matter of a current debate, and we experimentally establish a new microscopic scenario.

\begin{figure*}[t]
   \centerline{\includegraphics{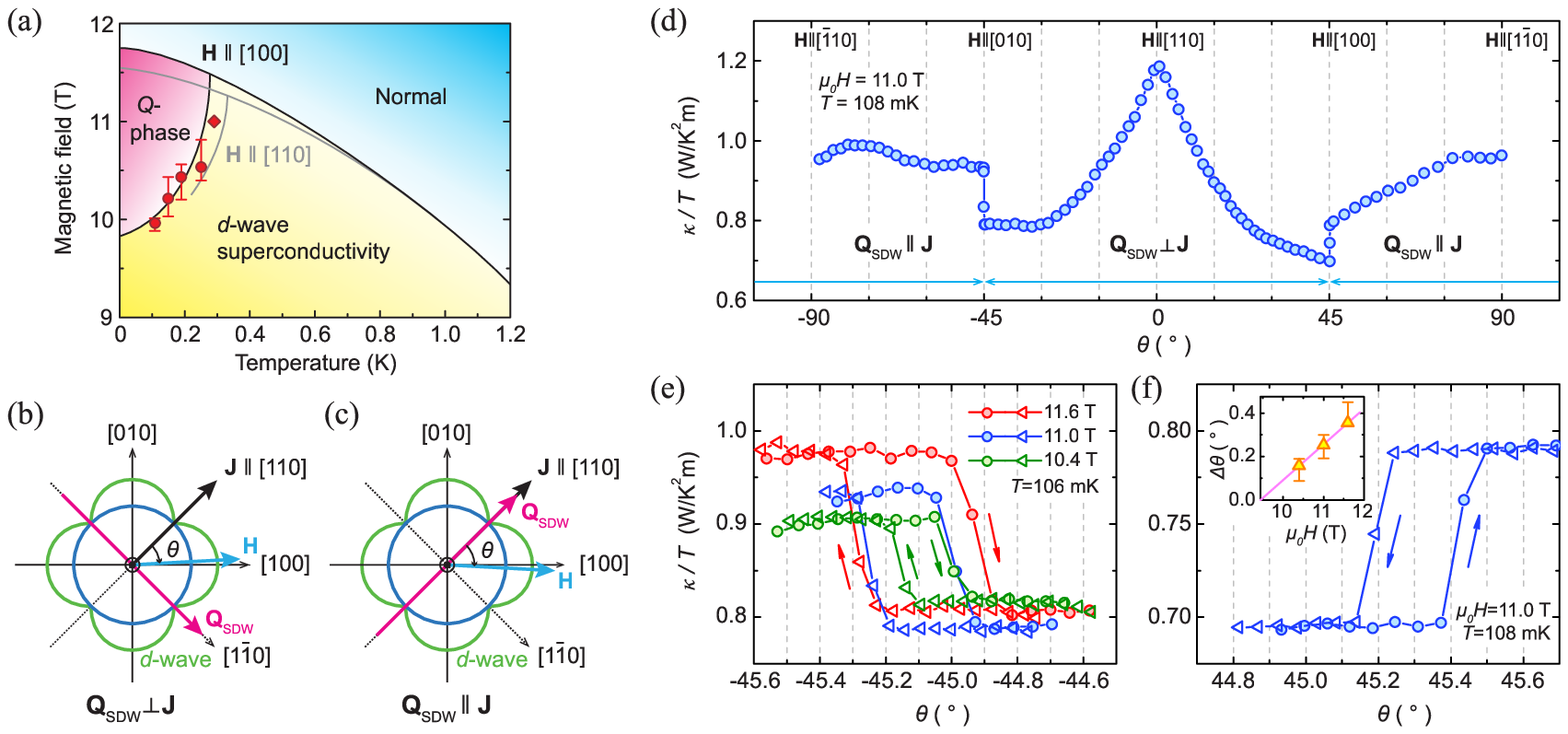}}
   \caption {(a) Phase diagram of CeCoIn$_5$, showing the $Q$ phase, based on specific heat measurements \cite{Bianchi2003}. The red data points are obtained from the present measurements, and the details are explained in Fig. 2. (b,c) Schematic diagrams that illustrate switching of the SDW magnetic domain ($\mathbf{Q}_\mathrm{SDW}$) as the magnetic field $\mathbf{H}$ is rotated about [100]. The heat current $\mathbf{J}$ is in the nodal [110] direction. $\mathbf{Q}_\mathrm{SDW}$ switches to be more perpendicular to $\mathbf{H}$, while lying along the nodes of the $d$-wave order parameter represented by the green curve. The blue circle represents the normal Fermi surface and the magnetization of the SDW points out of the plane. (d) The thermal conductivity of CeCoIn$_5$ $\kappa$ divided by temperature $T$ in the $Q$ phase as a function of the angle $\theta$ between $\mathbf{H}$ and the heat current $\mathbf{J}\parallel[110]$, at 11 T and 108 mK. The magnetic field is rotated between -90$^{\circ}$ and +90$^{\circ}$ within the crystallographic $ab$ plane. At 45$^{\circ}$ and -45$^{\circ}$, the antiferromagnetic ordering vector $\mathbf{Q}_\mathrm{SDW}$ switches between (0.44, 0.44, 0.5) and (0.44, -0.44, 0.5), as in (b,c) \cite{Gerber2014}. When $\mathbf{Q}_\mathrm{SDW}$ switches from $\mathbf{Q}_\mathrm{SDW}\perp\mathbf{J}$ to $\mathbf{Q}_\mathrm{SDW}\parallel\mathbf{J}$, the thermal conductivity increases by approximately 15\%. (e) Hysteretic behavior of the thermal conductivity in the switching region around $\theta=-45^{\circ}$, showing a first-order-like anomaly, for several fields. (f) Hysteretic behavior around $\theta=45^{\circ}$. The inset shows the width of the hysteresis as a function of magnetic field [from (e)].}
   \label{Fig1}
\end{figure*}

Though neutron scattering has been essential in identifying the nature of magnetism in the $Q$ phase, it does not probe the superconducting state with which magnetism couples. Thermal conductivity, however, is a powerful probe of superconductivity \cite{Matsuda2006,Shakeripour} because it depends on the presence of normal quasiparticles (excitations), as the superconducting condensate itself does not carry heat. Thermal conductivity is particularly sensitive to the presence of states where the energy gap in an unconventional superconductor is zero, i.e., gap nodes. This sensitivity arises because normal quasiparticles are easily excited around the nodes, where the energy gap is small, and therefore dominate the heat transport. As we show, measurements on the thermal conductivity of CeCoIn$_5$ in a rotating magnetic field reveal the nature of the $Q$ phase.

\section{Thermal conductivity in the $\bm{Q}$ phase}
\subsection{Experimental details}

 A needlelike single-crystal sample ($2.5\times 0.5\times 0.2~mm^3$) was prepared with the long axis along the [110] crystallographic direction that coincides with superconducting nodes. The heat current ($\mathbf{J}$) was applied along the [110] direction, and the thermal conductivity was measured with the standard steady-state method with two thermometers that were calibrated in advance. The magnetic field was applied within the crystallographic $ab$ plane, and the crystal (equivalently, magnetic field) was rotated about the $c$ axis using an Attocube piezoelectric rotator \cite{Yeoh}.
 
The alignment of the crystallographic axis was confirmed by Laue x-ray diffraction to be within 1$^\circ$. A total of eight sections (approximately 1 cm) of 50-$\mu$m-diameter platinum wire were spot welded to the sample, and small amounts of silver epoxy were applied over the welds for mechanical strength. The cold end of the sample was rigidly attached to a sample holder, a semicylindrical copper rod 2 mm in diameter. The sample was glued to the sample holder with varnish first; a pair of the platinum wires were wrapped around the sample and the sample holder; as the final step, silver paint was applied around the Pt wires, the sample holder, and the cold end of the sample, to enhance the electrical contacts between the bound wires and the sample holder and to ensure mechanical stability of the sample. The remaining three pairs of wires were used for thermal connections to two thermometers and a heater. The angle between the crystal and the magnetic field was monitored with two Hall sensors, parallel to the $ac$ and $bc$ planes, mounted on the sample stage.
 
\subsection{Results}

 As the magnetic field is rotated clockwise through [100] within the $ab$ plane, $\mathbf{Q}_\mathrm{SDW}$ flips from being perpendicular [Fig. 1(b)] to being parallel [Fig. 1(c)] to $\mathbf{J}$. Figure 1(d) shows the thermal conductivity as a function of the angle ($\theta$) between the magnetic field ($\mathbf{H}$) and the heat-current direction ($\mathbf{J}\parallel[110]$) at a temperature $T=108$ mK. The thermal conductivity exhibits sharp first-order jumps when the magnetic field is rotated around the [100] and [010] directions [Figs. 1(d)-1(f)] with a narrow hysteresis region of approximately 0.2$^\circ$ at 11 T. This response is identical to the switching of $\mathbf{Q}_\mathrm{SDW}$ observed by neutron diffraction \cite{Gerber2014} and, therefore, reflects the same hypersensitivity phenomenon. The width of the hysteresis is roughly linear with the magnetic field [Fig. 1(e) and inset of Fig. 1(f)] and tracks the development of the magnetic Bragg peak intensity \cite{Gerber2014}.
 
\begin{figure}[t]
   \centerline{\includegraphics[width=1\columnwidth]{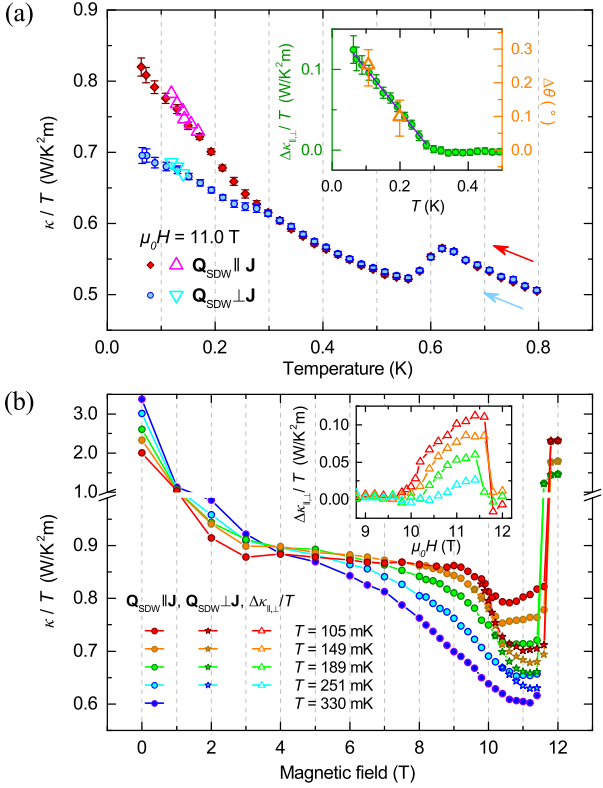}}
   \caption{(a) Temperature dependence of the thermal conductivity over $T$ ($\kappa/T$) for $\theta$=44.7$^\circ$ when $\mathbf{Q}_\mathrm{SDW}\perp\mathbf{J}$ (blue symbols), and for $\theta$=45.7$^\circ$ when $\mathbf{Q}_\mathrm{SDW}\parallel\mathbf{J}$ (red symbols) at a fixed field. The step at 0.6 K reflects the superconducting--normal transition. The data with higher thermal gradients (cyan and magenta triangles) show no difference from the other data ($\Delta T/T\approx 0.06$); i.e., there are no indications of the sliding mode of the SDW (Appendix D). Inset: The difference ($\Delta \kappa_{\parallel, \perp}/T$) between $\kappa/T$  for the two orientations of $\mathbf{Q}_\mathrm{SDW}$ (green circles, left axis) and the width of the hysteresis $\Delta \theta$ (from Fig. 6) at two temperatures for $\mu_0$H = 11 T (orange triangles, right axis). The onset temperature of the $Q$ phase at $\mu_0$H=11 T is depicted as a diamond on the phase diagram in Fig. 1(a). (b) Magnetic-field dependence of the thermal conductivity over $T$ ($\kappa/T$) at several temperatures. The field directions for the two different $\mathbf{Q}_\mathrm{SDW}$ orientations are the same as in (a). The inset shows the difference between $\kappa/T$ for the two orientations. The onset of the rise in $\Delta \kappa_{\parallel, \perp}/T$ is taken as a $Q$-phase transition and is displayed as red circles in Fig. 1(a).} 
   \label{Fig2}
\end{figure}

\begin{figure}[t]
   \centerline{\includegraphics[width=1\columnwidth]{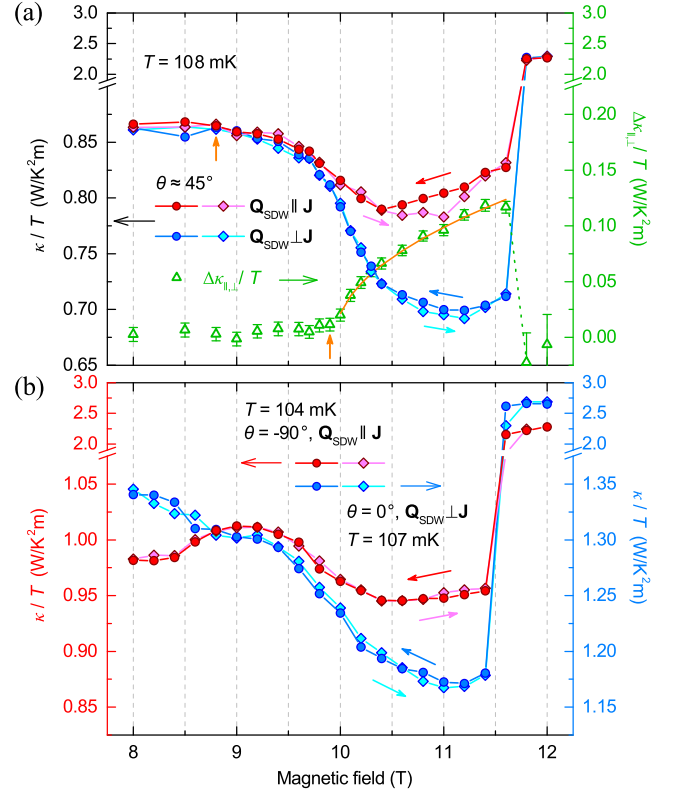}}
   \caption{(a) Magnetic-field dependence of the thermal conductivity over $T$ ($\kappa/T$) in and around the $Q$ phase. The field directions are the same as in Fig. 2. The data for field sweeps down (circles) and up (diamonds) are shown. The difference ($\Delta \kappa_{\parallel, \perp}/T$) for the two orientations of $\mathbf{Q}_\mathrm{SDW}$ (green triangles, right axis) starts to grow above 9.9 T, and is well described by the $\sqrt{(H-H_c)/H_c}$ fit shown (orange curve). Weak hysteresis within the $Q$ phase is likely due to a vortex-lattice transition for $\mathbf{H}\parallel[100]$ at $\mu_0$H$\approx$11 T, observed recently with scanning tunneling microscopy \cite{Feldman2016}. (b) $\kappa/T$ for $\theta$=-90$^\circ$ when $\mathbf{Q}_\mathrm{SDW}\parallel\mathbf{J}$ (red symbols), and $\theta$=0$^\circ$ when $\mathbf{Q}_\mathrm{SDW}\perp\mathbf{J}$ (blue symbols). The data are very similar to those for the same relative orientation of $\mathbf{Q}_\mathrm{SDW}$ and $\mathbf{J}$, correspondingly colored, in (a). The scales of the y axes in (a) and (b) are the same but with different offsets. The features for both $\theta$=-90$^\circ$and $\theta$=0$^\circ$ between 8 and 9 tesla are likely due to a vortex-lattice transition for $\mathbf{H}\parallel[110]$ between 7.5 and 8.7 T \cite{Das2012}. } 
   \label{Fig3}
\end{figure}

Figure 2 displays the thermal conductivity over temperature ($\kappa/T$) for two magnetic-field directions very close to the switching region at $\theta$=45$^\circ$, with $\mathbf{Q}_\mathrm{SDW}\parallel\mathbf{J}$ ($\theta$=45.7$^\circ$, red) and $\mathbf{Q}_\mathrm{SDW}\perp\mathbf{J}$ ($\theta$=44.7$^\circ$, blue). The temperature dependence at 11 T in Fig. 2(a) shows that the difference between the thermal conductivities for the two directions (see the inset) develops below 0.3 K. A similar increase of thermal conductivity with decreasing temperature at high fields was recently reported for $\mathbf{J}\parallel[100]$ \cite{Paglione2016}. 
The magnetic-field dependence in Fig. 2(b) also shows that the difference in $\kappa/T$ for the two directions of $\mathbf{Q}_\mathrm{SDW}$ develops at high magnetic field. The magnetic-field intensities of the onset of the increase of $\Delta \kappa/T$ from zero, with the corresponding measurement temperature, are displayed in the phase diagram of Fig. 1(a). These points coincide with the $Q$-phase boundary.

The temperature and field dependence of CeCoIn$_5$ is complex, and has not been reproduced in detail theoretically. The task of including (1) $d$-wave superconductivity, (2) a magnetic field which leads to both growth of the density of states due to the Doppler shift of the quasiparticle energies (the so-called Volovik effect) and a decrease of a quasiparticle mean-free path due to increased vortex scattering, (3) Pauli limiting, and (4) non-Fermi-liquid (NFL) behavior in the vicinity of H$_{c2}$, even leaving out the SDW of the $Q$ phase, is a monumental one. We can, however, offer potential explanations of some of the observed trends based on the phenomena mentioned above. For example, the increase of thermal conductivity with reducing temperature within the superconducting state in magnetic fields close to H$_{c2}$ [Fig. 2(a)] can be attributed to a similar NFL behavior in the normal state at or above H$_{c2}$ \cite{Ronning2005,Paglione2016}.

 As shown by the data for the lowest temperature of 105 mK in Fig. 2(b), the thermal conductivity is flat as a function of field between 4 and 9 tesla. Therefore, any deviation of $\kappa/T$ from the flat behavior in the high-field regime (above 9 T) should be attributed to the formation of the $Q$ phase. 

Figure 3 displays the thermal conductivity data inside and around the $Q$ phase. The abrupt changes at H$_{c2}$, 11.7 T for H$\parallel$[100] and 11.5 T for H$\parallel$[110] and H$\parallel[\bar{1}10]$, agree well with the first-order transitions found in previous studies \cite{Bianchi2003}. There are a couple of salient features: (1) The difference between $\kappa/T$ for the two orientations of $\mathbf{Q}_\mathrm{SDW}$ [right axis of Fig. 3(a)] grows above 9.9 T and drops abruptly to zero above H$_{c2}$, similar to the behavior of the SDW intensity measured by neutron scattering \cite{Gerber2014}. The functional dependence, however, is different: While neutron intensity $I\propto(H-H_c)$, $\Delta \kappa/T \propto \sqrt{H-H_c}$ as shown in Fig. 3(a) by the orange curve. The ordered magnetic moment $M\propto \sqrt{I} \propto \sqrt{H-H_c}$ \cite{Gerber2014}. Therefore, $\Delta \kappa/T$ grows linearly with the magnetic moment. (2) For both directions of $\mathbf{Q}_\mathrm{SDW}$, $\kappa/T$ starts to decrease around 9 T, well before entering the $Q$ phase. This reduction of $\kappa/T$ may be related to the results of NMR measurements \cite{Koutroulakis2010} that were interpreted as an additional phase, but it may also arise from fluctuations due to the quantum-critical point  ($T=0, H\approx9.8$ T)  associated with the $Q$ phase.

The key observations in current measurements are that, for both orientations of $\mathbf{Q}_\mathrm{SDW}$, the thermal conductivity drops in the $Q$ phase, and the reduction is larger for $\mathbf{Q}_\mathrm{SDW}\perp\mathbf{J}$. The data in Fig. 3 also show that $\kappa/T$ for  $\theta$=0$^\circ$ and 44.7$^\circ$, with $\mathbf{Q}_\mathrm{SDW}\perp\mathbf{J}$, closely reproduce each other as do the data for $\theta$=45.7$^\circ$ and -90$^\circ$ with $\mathbf{Q}_\mathrm{SDW}\parallel\mathbf{J}$. Whatever changes in the $Q$ phase to affect thermal conductivity as the magnetic field rotates, those changes occur abruptly at $\theta$=45$^\circ$ and then remain unchanged for the subsequent 90$^\circ$ of the field rotation.

\section{Discussion}

SDW order alone \cite{Mineev2015} cannot account for our data. SDW gaps quasiparticles along the $\mathbf{Q}_\mathrm{SDW}$, and as a result, the thermal conductivity along the $\mathbf{Q}_\mathrm{SDW}$ must be smaller than the thermal conductivity perpendicular to it, contrary to our observation. There must be an additional component in the $Q$ phase that has an opposite and stronger effect on thermal conductivity compared to that of its SDW.

A proposal for the origin of the hypersensitivity of $\mathbf{Q}_\mathrm{SDW}$ based on the formation of the FFLO state \cite{Hatakeyama2015} is also incompatible with our result. Within this theory, $\bm{q}_{\mathrm{FFLO}}\parallel\mathbf{H}$, which leads to a smooth change of  the effect of the FFLO state on the thermal conductivity as both $\mathbf{H}$ and $\bm{q}_{\mathrm{FFLO}}$ rotate together through [100]. Therefore, the influence of the SDW will dominate, and $\kappa/T(\mathbf{Q}_\mathrm{SDW}\parallel\mathbf{J}) < \kappa/T(\mathbf{Q}_\mathrm{SDW}\perp\mathbf{J}$) should be observed in the vicinity of $\theta$=45$^\circ$, in contrast to the experiment. To possibly reconcile this theory with the data, the requirement that $\bm{q}_{\mathrm{FFLO}}\parallel\mathbf{H}$ must be relaxed, with $\bm{q}_{\mathrm{FFLO}}$ pointing along the $d$-wave nodes (see Appendix C).

\begin{figure}[tb]
   \centerline{\includegraphics[width=1\columnwidth]{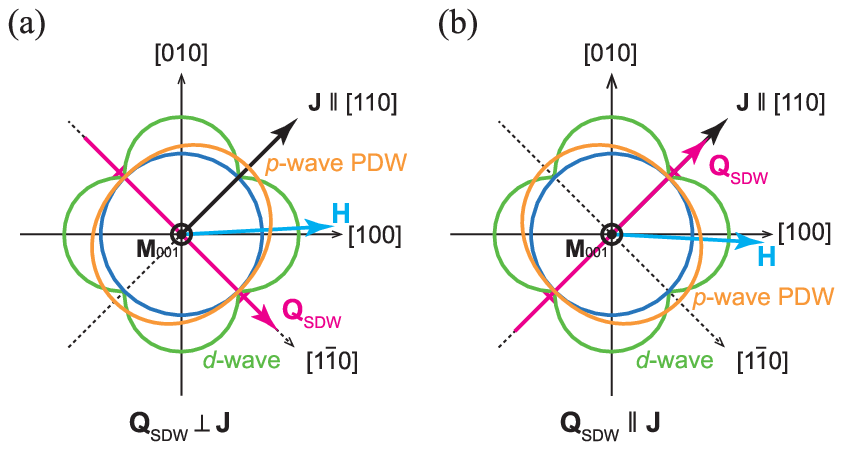}}
   \caption{Schematics of the $d$-wave (green) and the $p$-wave PDW (orange,  described by the vector $\mathbf{d}_1$) superconducting order parameters, the $\mathbf{Q}_\mathrm{SDW}$ vectors and the corresponding gaps (magenta arrow and arcs, respectively), and the magnetic-field directions (cyan arrow). A combination of a SDW \cite{Mineev2015} and a $p$-wave PDW \cite{Agterberg,Gerber2014} $\mathbf{d}_1$ component is consistent with the thermal conductivity data. (a) When $\mathbf{Q}_\mathrm{SDW}\perp\mathbf{J}$, the $p$-wave antinodes gap the nodes of the $d$ wave along $\mathbf{J}$ (black arrow), sharply reducing the thermal conductivity. (b) The effect of the SDW gapping the nodes along $\mathbf{J}$ must be smaller (by a factor of approximately 2) than the similar effect of the $p$-wave PDW state in (a).} 
   \label{Fig4}
\end{figure}

A natural explanation that accounts for the observed reduction of the thermal conductivity in the $Q$ phase is the existence of a spatially inhomogeneous $p$-wave PDW that couples the superconducting $d$-wave and SDW order parameters \cite{Agterberg,Gerber2014}. One of the two $p$-wave PDW components (Appendix A), compatible with $d$-wave and SDW order parameters in CeCoIn$_5$, is $\mathbf{d}_1(k) = (0, 0, k_x-k_y)$ for $\mathbf{Q}_\mathrm{SDW} \parallel[110]$ or, equivalently, $\mathbf{d}_1(k) = (0, 0, k_x+k_y)$ for $\mathbf{Q}_\mathrm{SDW}\parallel[1\bar{1}0]$,  shown schematically in Figs. 4(b) and 4(a), respectively. An interplay between the SDW and the PDW $\mathbf{d}_1$ locks the node of the $p$ wave along the direction of $\mathbf{Q}_\mathrm{SDW}$, leading to an additional anisotropy of the thermal conductivity. To allow the SDW with $\mathbf{Q}_\mathrm{SDW}\parallel[1\bar{1}0]$ to form, as illustrated in Fig. 4(a), $\mathbf{d}_1$ has to leave the quasiparticles along $\mathbf{Q}_\mathrm{SDW}$ ungapped by aligning its nodes along this direction, which is also a nodal direction of the $d$ wave. The $p$-wave antinodes then gap the remaining $d$-wave nodes along [110] and reduce the thermal conductivity for $\mathbf{Q}_\mathrm{SDW}\perp\mathbf{J}$ [Fig. 4(a)]. We estimate that the average amplitude of the $p$-wave gap required to suppress the thermal conductivity for $\mathbf{Q}_\mathrm{SDW}\perp\mathbf{J}$ by 19\%, as observed experimentally at 108 mK and 11 T [Fig. 3(a)], is approximately 20\% of the primary $d$-wave gap, and the magnitude of the SDW gap is comparatively smaller and approximately 10\% of the $d$-wave gap (see Appendix B).

A hierarchy of interactions between various orders shown in Fig. 4 accounts for both the hypersensitivity and the thermal conductivity data. (1) The SDW must lie along one of the nodes of the superconducting $d$-wave order parameter. (2) The spin-orbit coupling effect on the interaction between the SDW and the magnetic field drives the hypersensitivity \cite{Mineev2015} and orients $\mathbf{Q}_\mathrm{SDW}$ as perpendicular to $\mathbf{H}$ as possible. (3) The selected $\mathbf{Q}_\mathrm{SDW}$, in turn, orients the (allowed) $p$-wave PDW $\mathbf{d}_1$ component. Finally, (4) the PDW gaps the $d$-wave nodes more effectively than the SDW, leading to the observed trend in thermal conductivity.

Our measurements demonstrate a macroscopic realization of intertwined orders. As systems with multiple orders are becoming increasingly common in correlated electronic materials, we expect more examples of similar intertwined orders in which the manipulation of one order by the other is possible.

\begin{acknowledgments}
Discussions with James A. Sauls, Anton B. Vorontsov, Ilya Vekhter, Stuart E. Brown, Alexander V. Balatsky, David M. Fobes, and Marc Janoschek are gratefully acknowledged. This work was conducted at the Los Alamos National Laboratory under the auspices of the U.S. Department of Energy, Office of Basic Energy Sciences, Division of Materials Sciences and Engineering. We gratefully acknowledge the support of the U.S. Department of Energy through the LANL/LDRD Program.
\end{acknowledgments}

\appendix

\section{Scenario for the Hypersensitivity of $\mathbf{Q}_\mathrm{SDW}$ on the Direction of the Magnetic Field based on the \textit{p}-wave PDW}

The switching of $\mathbf{Q}_\mathrm{SDW}$, reported in Ref. \cite{Gerber2014}, was suggested by the authors to be due to the formation of the spatially inhomogeneous $p$-wave PDW, which, in a $d$-wave superconductor, couples to the SDW \cite{Agterberg}. It was suggested that the anisotropic magnetic susceptibility of the $p$-wave component orients it with respect to the magnetic field, and the interaction between the PDW and the SDW will then orient $\mathbf{Q}_\mathrm{SDW}$. The two $p$-wave components within the PDW scenario, compatible with $d$-wave and SDW order parameters in CeCoIn$_5$, are $\mathbf{d}_1(\mathbf{k})=(0,0,k_x-k_y)$ and $\mathbf{d}_2(\mathbf{k})=(k_z, -k_z, 0)$  for $\mathbf{Q}_\mathrm{SDW}\parallel[110]$ \cite{Gerber2014,Agterberg}. Magnetic susceptibility of the $\mathbf{d}_2$ component is indeed anisotropic. Its nodal plane, however, lies within the $ab$ plane, and $\mathbf{d}_2$, therefore, cannot preferentially select one of the two possible $\mathbf{Q}_\mathrm{SDW}$. The $\mathbf{d}_1$ component has nodes along [110] and would select the SDW domain with $\mathbf{Q}_\mathrm{SDW}$ along that direction. However, $\mathbf{d}_1$ has isotropic $ab$-plane susceptibility, and it cannot be the sole source of the hypersensitivity of $\mathbf{Q}_\mathrm{SDW}$. The $p$-wave PDW, therefore, can be the cause of the hypersensitivity of $\mathbf{Q}_\mathrm{SDW}$ only when the $\mathbf{d}_1(\mathbf{k})$ and $\mathbf{d}_2(\mathbf{k})$ are coupled. Currently, there is no theoretical support for the existence of such a coupling in CeCoIn$_5$. Consequently, we exclude the possibility that the anisotropy of the magnetic susceptibility of $\mathbf{d}_2$ is the origin of the field hypersensitivity. Nevertheless, the triplet $\mathbf{d}_1$ component directly couples to the SDW and $d$-wave orders, it is allowed to form in the $Q$ phase, and it does explain our thermal conductivity results.

\section{Contribution of the Composite Order Parameter to the Thermal Conductivity}

To ascertain which nodes of the $d$-wave order parameter contribute most to heat transport, we calculated the thermal conductivity using the theory in Ref. \cite{Vorontsov2007} for a superconductor with a composite order parameter, $|\Delta_d|+i~a|\Delta_p|$, where $|\Delta_d|$ and $|\Delta_p|$ are the magnitudes of the $d$-wave and the $p$-wave ($\mathbf{d}_1$) components, respectively. The heat current $\mathbf{J}$ was taken to be along one of the nodes of the $d$-wave gap. The nodes of the $p$-wave order parameter were arranged to either coincide with $\mathbf{J}$ (and one of the nodal directions of the $d$ wave) or to be perpendicular to it. Addition of the $p$-wave component with imaginary phase guaranteed that the antinodes of the $p$ wave gap the nodes of the $d$ wave parallel to them and reduce the thermal conductivity of their quasiparticles. The calculations, shown in Fig. 5, demonstrate that thermal conductivity is reduced much more, between a factor of 5 and 10, when the $p$-wave nodes are perpendicular to $\mathbf{J}$; i.e., the $p$-wave antinodes gap the $d$-wave nodes that are along the heat transport. This means that the $d$-wave nodes along the heat flow dominate thermal transport because the velocity of the quasiparticles in these nodes has a large component parallel to the direction of the heat current.

\begin{figure}[t]
   \centerline{\includegraphics[width=1\columnwidth]{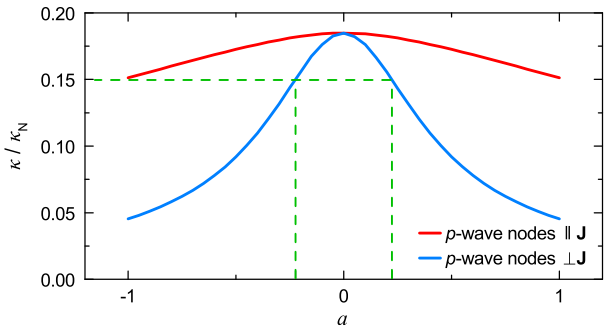}}
   \caption {The calculated thermal conductivity as a function of the relative amplitude $a$ of the $p$-wave component for $|\Delta_d|+i~a|\Delta_p|$ pairing symmetry without normalization for two orientations of the $p$-wave component $\mathbf{d}_1$ with respect to the heat current $\mathbf{J}$ depicted in Figs. 4(a) and 4(b) of the main text. The nodes of the $p$-wave component are either perpendicular to $\mathbf{J}$ [Fig. 4(a)] or parallel to it [Fig. 4(b)]. The electron mean-free path $l=10\xi$, where $\xi$ is the superconducting coherence length, $T=0.05T_c$, and $H=0.3H_{c2}$, where $H_{c2}$ is the orbital upper critical field. The reduction of $\kappa$ is much stronger when the $p$-wave antinode is along the heat current ($p$-wave $\mathrm{nodes}\perp\mathbf{J}$) because the $p$-wave antinode gaps the $d$-wave nodal quasiparticles with momenta along the heat current $\mathbf{J}$. In contrast, when the $p$-wave antinode is perpendicular to $\mathbf{J}$ ($p$-wave $\mathrm{nodes}\parallel\mathbf{J}$), the $p$-wave antinode only gaps quasiparticles with momenta perpendicular to the heat current, resulting in a much smaller effect. The thermal transport in a $d$-wave superconductor is therefore dominated by the quasiparticles in the nodes that are along the heat current. }
   \label{Fig5}
\end{figure}

 These calculations also allow us to estimate the relative magnitude ($a$) of the $p$-wave order parameter required to achieve the reduction of thermal conductivity by 19\% observed experimentally for the case of $\mathbf{Q}_\mathrm{SDW}\perp\mathbf{J}$. The hypersensitivity of $\mathbf{Q}_\mathrm{SDW}$ is due to the spin-orbit coupling, whereas, the observed anisotropic thermal conductivity is due to the appearance of the allowed $\mathbf{d}_1$ component of the $p$-wave PDW. The reduction of thermal conductivity when $\mathbf{Q}_\mathrm{SDW}$ points along the heat current $\mathbf{J}$ ($\mathbf{Q}_\mathrm{SDW}\parallel\mathbf{J}$, when the dominant nodes along $\mathbf{J}$ are gapped by the SDW) is less than half of the reduction for the case of $\mathbf{Q}_\mathrm{SDW}\perp\mathbf{J}$, where the nodes along $\mathbf{J}$ are gapped by the secondary $p$-wave component. The contribution of the SDW to the reduction of thermal conductivity in the latter case is reduced even further, by a factor of 10, as seen in Fig. 5. Therefore, when $\mathbf{Q}_\mathrm{SDW}\perp\mathbf{J}$, we can neglect the effect of SDW on the thermal conductivity for the purpose of making an estimate of the magnitude of the $p$-wave order parameter. We then consider the case of the $p$-wave nodes perpendicular to the heat current, shown in Fig. 5. The horizontal dashed line represents the observed suppression of thermal conductivity for $\mathbf{Q}_\mathrm{SDW}\perp\mathbf{J}$, and we can read off the magnitude of the $p$-wave $\mathbf{d}_1$ component from their intersections with the blue ($p$-wave $\mathrm{nodes}\perp\mathbf{J}$) curve (vertical dashed lines). The resulting $a\approx0.2$, a reasonable number for the amplitude of a secondary superconducting order parameter. We can roughly estimate the magnitude of the SDW gap (or the equivalent $p$-wave gap) required to suppress the thermal conductivity by 8\%, as observed experimentally for $\mathbf{Q}_\mathrm{SDW}\parallel\mathbf{J}$. We obtain the SDW gap to be approximately 10\% of the primary $d$-wave order parameter, half of the average $p$-wave PDW gap.
 
\begin{figure}[t]
   \centerline{\includegraphics[width=1\columnwidth]{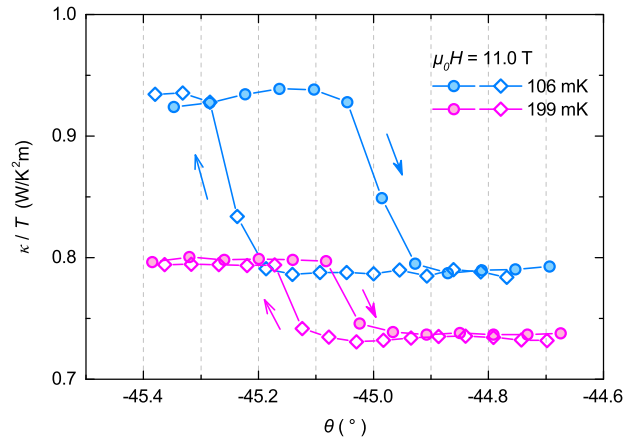}}
   \caption{The hysteresis of the hypersensitive switching via thermal conductivity around $\theta=-45^{\circ}$ at 11 T and two temperatures, 106 and 199 mK. The data for 106 mK are the same as in Fig. 1(e) of the main text. The widths of the hysteresis at two different temperatures are plotted in the inset of Fig. 2(a). } 
   \label{Fig6}
\end{figure}

\section{FFLO state as the Origin of Hypersensitivity vis-\`a-vis Thermal Conductivity in the $\bm{Q}$ phase}

Strong Pauli-limiting effects \cite{Maki}, evidenced by a first-order superconducting transition \cite{Bianchi2003} above 10 T, and an extremely long electron mean-free path in the superconducting state \cite{Movshovich2001} raise the possibility of the formation of a spatially inhomogeneous FFLO state. The FFLO state is characterized by a wave vector $\bm{q}_{\mathrm{FFLO}}$, with the superconducting order parameter in the Larkin-Ovchinnikov (LO) scenario varying as $\Delta=|\Delta|\mathrm{cos}(\bm{q}_{\mathrm{FFLO}}\cdot \bm{r})$ and leading to a periodic array of nodal planes perpendicular to $\bm{q}_{\mathrm{FFLO}}$ where the superconducting gap ($\Delta$) is zero. There is a number of experiments that are consistent with a FFLO state in CeCoIn$_5$. One of the most notable works is the NMR investigation \cite{kumagai2011} that showed the resonance signal expected from the normal electrons in the FFLO nodal planes. The fragile nature of the $Q$ phase found in the doping experiment \cite{Tokiwa2010} also implies the existence of the FFLO state \cite{Ikeda2010}.

\begin{figure}[t]
   \centerline{\includegraphics[width=1\columnwidth]{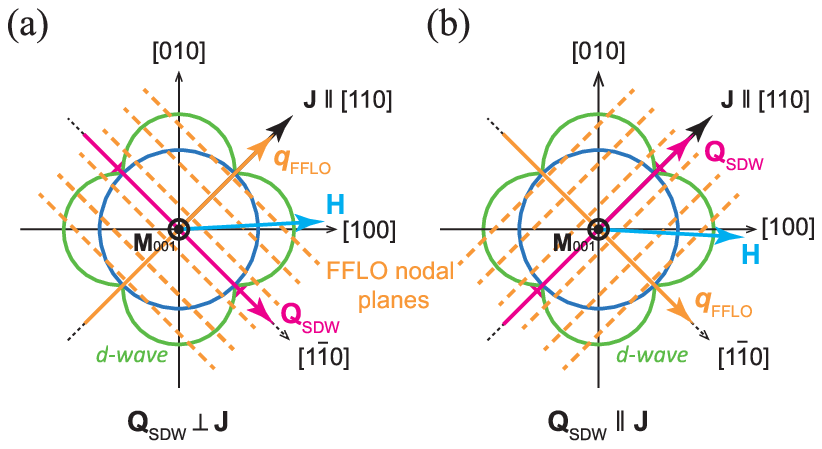}}
   \caption{Possible configuration of a FFLO state in CeCoIn$_5$. Schematic of the $d$-wave and the FFLO order parameters, the $\mathbf{Q}_\mathrm{SDW}$ vector, and the magnetic-field directions that are compatible with both hypersensitivity and thermal conductivity within a modified version of the FFLO scenario \cite{Hatakeyama2015}. This modification would require that  $\bm{q}_{\mathrm{FFLO}}$ be forced to lie along the nodes of the $d$ wave, instead of always being parallel to the applied magnetic field. Dashed lines indicate the nodes of the FFLO state with normal quasiparticles. (\textbf{a}) FFLO nodal planes increase scattering and reduce thermal conductivity when they are perpendicular to $\mathbf{J}$. (\textbf{b}) The nodal planes contribute to thermal transport along $\mathbf{J}$ when they are parallel to it.} 
   \label{Fig7}
\end{figure}

As stated in the main text, the proposal for the origin of the hypersensitivity of $\mathbf{Q}_\mathrm{SDW}$ based on the formation of a FFLO state \cite{Hatakeyama2015} does not explain the thermal conductivity data. Within this theory, $\bm{q}_{\mathrm{FFLO}}$ is parallel to $\mathbf{H}$, and $\bm{q}_{\mathrm{FFLO}}$ rotates gradually through [100] together with the magnetic field. The FFLO state will therefore provide a smooth background to the thermal conductivity as the field rotates. With only a SDW present in addition to a FFLO state, the SDW will dominate the response in the vicinity of the switching region around $\theta=45^{\circ}$ and give $\kappa/T(\mathbf{Q}_\mathrm{SDW}\parallel\mathbf{J})<\kappa/T(\mathbf{Q}_\mathrm{SDW}\perp\mathbf{J})$, which is in contrast to the experimental result. To reconcile this theory with the data, the requirement that $\bm{q}_{\mathrm{FFLO}}\parallel\mathbf{H}$ must be relaxed.

In fact, $\bm{q}_{\mathrm{FFLO}}$ was shown to lie along the nodes in the majority of the FFLO phase or along the antinodes of the $d$-wave order parameter \cite{Vorontsov2005} when orbital (vortex) effects were not considered. The interaction between a FFLO and a SDW \cite{Hatakeyama2015,Yanase2011,Bulaevskii} also prefers $\mathbf{Q}_\mathrm{SDW}\perp\bm{q}_{\mathrm{FFLO}}$ and would therefore tend to allign $\bm{q}_{\mathrm{FFLO}}$ along the $d$-wave nodes in the $Q$ phase of CeCoIn$_5$. This alignment occurs because Cooper pairs traveling in a direction perpendicular to $\mathbf{Q}_\mathrm{SDW}$ experience a uniform magnetization, and it is preferable for superconductivity to be modulated in this direction \cite{Bulaevskii}. If these requirements were allowed to be satisfied, i.e., if $\bm{q}_{\mathrm{FFLO}}$ is allowed to not follow $\mathbf{H}$ and to instead lie along the $d$-wave nodes and be perpendicular to $\mathbf{Q}_\mathrm{SDW}$, the following will take place: For $\mathbf{Q}_\mathrm{SDW}\perp\mathbf{J}$, $\bm{q}_{\mathrm{FFLO}}\parallel \mathbf{J}$, and the FFLO nodal planes would be perpendicular to $\mathbf{J}$ [Fig. 7(a)] and increase quasiparticle scattering, decreasing $\kappa$. For $\mathbf{Q}_\mathrm{SDW}\parallel\mathbf{J}$, $\bm{q}_{\mathrm{FFLO}}\perp\mathbf{J}$, and the FFLO nodal planes would be parallel to $\mathbf{J}$ [Fig. 7(b)], increasing both the density of states of quasiparticles with momentum $\bm{k}$ along $\mathbf{J}$ and $\kappa$. The effect of the FFLO state described therefore has the right trend and, if larger than the effect of the SDW, could explain the thermal conductivity data. The requirement that $\bm{q}_{\mathrm{FFLO}}$ cannot be allowed to follow the magnetic field is necessitated by the fact that the changes must take place abruptly at $\theta=45^{\circ}$, and after that the orders relevant to thermal conductivity must remain constant until the next antinodal plane of the $d$ wave (at $\theta=-45^{\circ}$ or $\theta=135^{\circ}$) is crossed by the applied magnetic field. 

In summary, to be compatible with our thermal conductivity data, the FFLO-based scenario for hypersensitivity \cite{Hatakeyama2015} must be modified to allow $\bm{q}_{\mathrm{FFLO}}$ to be locked to the nodal direction of the primary $d$-wave order parameter, with a possibility that needs to be tested theoretically and experimentally. In particular, small-angle neutron-scattering (SANS) measurements, with the neutron flux along the nodal [110] direction and the magnetic field applied to select $\bm{q}_{\mathrm{FFLO}}\parallel[1\bar{1}0]$, may reveal the FFLO state if it is present.

\section{Sliding Mode of a Spin-Density-Wave}

We rule out a contribution of the sliding mode of the SDW. A sliding mode along the ordering wave vector $\mathbf{Q}_\mathrm{SDW}$ can be expected in an incommensurate SDW state \cite{Lee}. Such effects depend heavily on pinning the SDW at impurity centers. When an incommensurate charge-density-wave (CDW) is depinned at a critical driving potential, the current it carries is a nonlinear function of the driving potential. We therefore expect a nonlinear response of thermal conductivity as a function of a sufficiently large thermal gradient in the sample. Our measurements for high thermal gradients (between 30\% and 80\% higher than normal) are also displayed in Fig. 2(a). We did not observe any changes in thermal conductivity as a function of thermal gradient. Either the SDW remains pinned by impurities, or the contribution of the sliding mode to thermal conductivity is negligible.

\bibliography{QPhase}

\begin{thebibliography}{46}%
\makeatletter
\providecommand \@ifxundefined [1]{%
 \@ifx{#1\undefined}
}%
\providecommand \@ifnum [1]{%
 \ifnum #1\expandafter \@firstoftwo
 \else \expandafter \@secondoftwo
 \fi
}%
\providecommand \@ifx [1]{%
 \ifx #1\expandafter \@firstoftwo
 \else \expandafter \@secondoftwo
 \fi
}%
\providecommand \natexlab [1]{#1}%
\providecommand \enquote  [1]{``#1''}%
\providecommand \bibnamefont  [1]{#1}%
\providecommand \bibfnamefont [1]{#1}%
\providecommand \citenamefont [1]{#1}%
\providecommand \href@noop [0]{\@secondoftwo}%
\providecommand \href [0]{\begingroup \@sanitize@url \@href}%
\providecommand \@href[1]{\@@startlink{#1}\@@href}%
\providecommand \@@href[1]{\endgroup#1\@@endlink}%
\providecommand \@sanitize@url [0]{\catcode `\\12\catcode `\$12\catcode
  `\&12\catcode `\#12\catcode `\^12\catcode `\_12\catcode `\%12\relax}%
\providecommand \@@startlink[1]{}%
\providecommand \@@endlink[0]{}%
\providecommand \url  [0]{\begingroup\@sanitize@url \@url }%
\providecommand \@url [1]{\endgroup\@href {#1}{\urlprefix }}%
\providecommand \urlprefix  [0]{URL }%
\providecommand \Eprint [0]{\href }%
\providecommand \doibase [0]{http://dx.doi.org/}%
\providecommand \selectlanguage [0]{\@gobble}%
\providecommand \bibinfo  [0]{\@secondoftwo}%
\providecommand \bibfield  [0]{\@secondoftwo}%
\providecommand \translation [1]{[#1]}%
\providecommand \BibitemOpen [0]{}%
\providecommand \bibitemStop [0]{}%
\providecommand \bibitemNoStop [0]{.\EOS\space}%
\providecommand \EOS [0]{\spacefactor3000\relax}%
\providecommand \BibitemShut  [1]{\csname bibitem#1\endcsname}%
\let\auto@bib@innerbib\@empty
%</preamble>
\bibitem [{\citenamefont {Bardeen}\ \emph {et~al.}(1957)\citenamefont
  {Bardeen}, \citenamefont {Cooper},\ and\ \citenamefont {Schrieffer}}]{BCS}%
  \BibitemOpen
  \bibfield  {author} {\bibinfo {author} {\bibfnamefont {J.}~\bibnamefont
  {Bardeen}}, \bibinfo {author} {\bibfnamefont {L.~N.}\ \bibnamefont {Cooper}},
  \ and\ \bibinfo {author} {\bibfnamefont {J.~R.}\ \bibnamefont {Schrieffer}},\
  }\href {\doibase 10.1103/PhysRev.106.162} {\bibfield  {journal} {\bibinfo
  {journal} {Phys. Rev.}\ }\textbf {\bibinfo {volume} {106}},\ \bibinfo {pages}
  {162} (\bibinfo {year} {1957})}\BibitemShut {NoStop}%
\bibitem [{\citenamefont {Werthamer}\ \emph {et~al.}(1966)\citenamefont
  {Werthamer}, \citenamefont {Helfand},\ and\ \citenamefont
  {Hohenberg}}]{Hohenberg}%
  \BibitemOpen
  \bibfield  {author} {\bibinfo {author} {\bibfnamefont {N.~R.}\ \bibnamefont
  {Werthamer}}, \bibinfo {author} {\bibfnamefont {E.}~\bibnamefont {Helfand}},
  \ and\ \bibinfo {author} {\bibfnamefont {P.~C.}\ \bibnamefont {Hohenberg}},\
  }\href {\doibase 10.1103/PhysRev.147.295} {\bibfield  {journal} {\bibinfo
  {journal} {Phys. Rev.}\ }\textbf {\bibinfo {volume} {147}},\ \bibinfo {pages}
  {295} (\bibinfo {year} {1966})}\BibitemShut {NoStop}%
\bibitem [{\citenamefont {Clogston}(1962)}]{Clogston}%
  \BibitemOpen
  \bibfield  {author} {\bibinfo {author} {\bibfnamefont {A.~M.}\ \bibnamefont
  {Clogston}},\ }\href {\doibase 10.1103/PhysRevLett.9.266} {\bibfield
  {journal} {\bibinfo  {journal} {Phys. Rev. Lett.}\ }\textbf {\bibinfo
  {volume} {9}},\ \bibinfo {pages} {266} (\bibinfo {year} {1962})}\BibitemShut
  {NoStop}%
\bibitem [{\citenamefont {Lake}\ \emph {et~al.}(2002)\citenamefont {Lake},
  \citenamefont {R{\o}nnow}, \citenamefont {Christensen}, \citenamefont
  {Aeppli}, \citenamefont {Lefmann}, \citenamefont {McMorrow}, \citenamefont
  {Vorderwisch}, \citenamefont {Smeibidl}, \citenamefont {Mangkorntong},
  \citenamefont {Sasagawa}, \citenamefont {Nohara}, \citenamefont {Takagi},\
  and\ \citenamefont {Mason}}]{Lake2002}%
  \BibitemOpen
  \bibfield  {author} {\bibinfo {author} {\bibfnamefont {B.}~\bibnamefont
  {Lake}}, \bibinfo {author} {\bibfnamefont {H.~M.}\ \bibnamefont {R{\o}nnow}},
  \bibinfo {author} {\bibfnamefont {N.~B.}\ \bibnamefont {Christensen}},
  \bibinfo {author} {\bibfnamefont {G.}~\bibnamefont {Aeppli}}, \bibinfo
  {author} {\bibfnamefont {K.}~\bibnamefont {Lefmann}}, \bibinfo {author}
  {\bibfnamefont {D.~F.}\ \bibnamefont {McMorrow}}, \bibinfo {author}
  {\bibfnamefont {P.}~\bibnamefont {Vorderwisch}}, \bibinfo {author}
  {\bibfnamefont {P.}~\bibnamefont {Smeibidl}}, \bibinfo {author}
  {\bibfnamefont {N.}~\bibnamefont {Mangkorntong}}, \bibinfo {author}
  {\bibfnamefont {T.}~\bibnamefont {Sasagawa}}, \bibinfo {author}
  {\bibfnamefont {M.}~\bibnamefont {Nohara}}, \bibinfo {author} {\bibfnamefont
  {H.}~\bibnamefont {Takagi}}, \ and\ \bibinfo {author} {\bibfnamefont {T.~E.}\
  \bibnamefont {Mason}},\ }\href {\doibase 10.1038/415299a} {\bibfield
  {journal} {\bibinfo  {journal} {Nature}\ }\textbf {\bibinfo {volume} {415}},\
  \bibinfo {pages} {299} (\bibinfo {year} {2002})}\BibitemShut {NoStop}%
\bibitem [{\citenamefont {Cai}\ \emph {et~al.}(2013)\citenamefont {Cai},
  \citenamefont {Zhou}, \citenamefont {Ruan}, \citenamefont {Wang},
  \citenamefont {Chen}, \citenamefont {Lee},\ and\ \citenamefont
  {Wang}}]{Cai2013}%
  \BibitemOpen
  \bibfield  {author} {\bibinfo {author} {\bibfnamefont {P.}~\bibnamefont
  {Cai}}, \bibinfo {author} {\bibfnamefont {X.}~\bibnamefont {Zhou}}, \bibinfo
  {author} {\bibfnamefont {W.}~\bibnamefont {Ruan}}, \bibinfo {author}
  {\bibfnamefont {A.}~\bibnamefont {Wang}}, \bibinfo {author} {\bibfnamefont
  {X.}~\bibnamefont {Chen}}, \bibinfo {author} {\bibfnamefont {D.-H.}\
  \bibnamefont {Lee}}, \ and\ \bibinfo {author} {\bibfnamefont
  {Y.}~\bibnamefont {Wang}},\ }\href {http://dx.doi.org/10.1038/ncomms2592}
  {\bibfield  {journal} {\bibinfo  {journal} {Nat. Commun.}\ }\textbf {\bibinfo
  {volume} {4}},\ \bibinfo {pages} {1596} (\bibinfo {year} {2013})}\BibitemShut
  {NoStop}%
\bibitem [{\citenamefont {Bianchi}\ \emph {et~al.}(2002)\citenamefont
  {Bianchi}, \citenamefont {Movshovich}, \citenamefont {Oeschler},
  \citenamefont {Gegenwart}, \citenamefont {Steglich}, \citenamefont
  {Thompson}, \citenamefont {Pagliuso},\ and\ \citenamefont
  {Sarrao}}]{Bianchi2002}%
  \BibitemOpen
  \bibfield  {author} {\bibinfo {author} {\bibfnamefont {A.}~\bibnamefont
  {Bianchi}}, \bibinfo {author} {\bibfnamefont {R.}~\bibnamefont {Movshovich}},
  \bibinfo {author} {\bibfnamefont {N.}~\bibnamefont {Oeschler}}, \bibinfo
  {author} {\bibfnamefont {P.}~\bibnamefont {Gegenwart}}, \bibinfo {author}
  {\bibfnamefont {F.}~\bibnamefont {Steglich}}, \bibinfo {author}
  {\bibfnamefont {J.~D.}\ \bibnamefont {Thompson}}, \bibinfo {author}
  {\bibfnamefont {P.~G.}\ \bibnamefont {Pagliuso}}, \ and\ \bibinfo {author}
  {\bibfnamefont {J.~L.}\ \bibnamefont {Sarrao}},\ }\href {\doibase
  10.1103/PhysRevLett.89.137002} {\bibfield  {journal} {\bibinfo  {journal}
  {Phys. Rev. Lett.}\ }\textbf {\bibinfo {volume} {89}},\ \bibinfo {pages}
  {137002} (\bibinfo {year} {2002})}\BibitemShut {NoStop}%
\bibitem [{\citenamefont {Bianchi}\ \emph {et~al.}(2003)\citenamefont
  {Bianchi}, \citenamefont {Movshovich}, \citenamefont {Capan}, \citenamefont
  {Pagliuso},\ and\ \citenamefont {Sarrao}}]{Bianchi2003}%
  \BibitemOpen
  \bibfield  {author} {\bibinfo {author} {\bibfnamefont {A.}~\bibnamefont
  {Bianchi}}, \bibinfo {author} {\bibfnamefont {R.}~\bibnamefont {Movshovich}},
  \bibinfo {author} {\bibfnamefont {C.}~\bibnamefont {Capan}}, \bibinfo
  {author} {\bibfnamefont {P.~G.}\ \bibnamefont {Pagliuso}}, \ and\ \bibinfo
  {author} {\bibfnamefont {J.~L.}\ \bibnamefont {Sarrao}},\ }\href {\doibase
  10.1103/PhysRevLett.91.187004} {\bibfield  {journal} {\bibinfo  {journal}
  {Phys. Rev. Lett.}\ }\textbf {\bibinfo {volume} {91}},\ \bibinfo {pages}
  {187004} (\bibinfo {year} {2003})}\BibitemShut {NoStop}%
\bibitem [{\citenamefont {Radovan}\ \emph {et~al.}(2003)\citenamefont
  {Radovan}, \citenamefont {Fortune}, \citenamefont {Murphy}, \citenamefont
  {Hannahs}, \citenamefont {Palm}, \citenamefont {Tozer},\ and\ \citenamefont
  {Hall}}]{Radovan2003}%
  \BibitemOpen
  \bibfield  {author} {\bibinfo {author} {\bibfnamefont {H.~A.}\ \bibnamefont
  {Radovan}}, \bibinfo {author} {\bibfnamefont {N.~A.}\ \bibnamefont
  {Fortune}}, \bibinfo {author} {\bibfnamefont {T.~P.}\ \bibnamefont {Murphy}},
  \bibinfo {author} {\bibfnamefont {S.~T.}\ \bibnamefont {Hannahs}}, \bibinfo
  {author} {\bibfnamefont {E.~C.}\ \bibnamefont {Palm}}, \bibinfo {author}
  {\bibfnamefont {S.~W.}\ \bibnamefont {Tozer}}, \ and\ \bibinfo {author}
  {\bibfnamefont {D.}~\bibnamefont {Hall}},\ }\href
  {http://dx.doi.org/10.1038/nature01842
  http://www.nature.com/nature/journal/v425/n6953/suppinfo/nature01842_S1.html}
  {\bibfield  {journal} {\bibinfo  {journal} {Nature}\ }\textbf {\bibinfo
  {volume} {425}},\ \bibinfo {pages} {51} (\bibinfo {year} {2003})}\BibitemShut
  {NoStop}%
\bibitem [{\citenamefont {Fulde}\ and\ \citenamefont {Ferrell}(1964)}]{FF}%
  \BibitemOpen
  \bibfield  {author} {\bibinfo {author} {\bibfnamefont {P.}~\bibnamefont
  {Fulde}}\ and\ \bibinfo {author} {\bibfnamefont {R.~A.}\ \bibnamefont
  {Ferrell}},\ }\href {\doibase 10.1103/PhysRev.135.A550} {\bibfield  {journal}
  {\bibinfo  {journal} {Phys. Rev.}\ }\textbf {\bibinfo {volume} {135}},\
  \bibinfo {pages} {A550} (\bibinfo {year} {1964})}\BibitemShut {NoStop}%
\bibitem [{\citenamefont {Larkin}\ and\ \citenamefont
  {Ovchinnikov}(1965)}]{LO}%
  \BibitemOpen
  \bibfield  {author} {\bibinfo {author} {\bibfnamefont {A.~I.}\ \bibnamefont
  {Larkin}}\ and\ \bibinfo {author} {\bibfnamefont {Y.~N.}\ \bibnamefont
  {Ovchinnikov}},\ }\href@noop {} {\bibfield  {journal} {\bibinfo  {journal}
  {Sov. Phys. JETP}\ }\textbf {\bibinfo {volume} {20}},\ \bibinfo {pages} {762}
  (\bibinfo {year} {1965})}\BibitemShut {NoStop}%
\bibitem [{\citenamefont {Young}\ \emph {et~al.}(2007)\citenamefont {Young},
  \citenamefont {Urbano}, \citenamefont {Curro}, \citenamefont {Thompson},
  \citenamefont {Sarrao}, \citenamefont {Vorontsov},\ and\ \citenamefont
  {Graf}}]{Young2007}%
  \BibitemOpen
  \bibfield  {author} {\bibinfo {author} {\bibfnamefont {B.-L.}\ \bibnamefont
  {Young}}, \bibinfo {author} {\bibfnamefont {R.~R.}\ \bibnamefont {Urbano}},
  \bibinfo {author} {\bibfnamefont {N.~J.}\ \bibnamefont {Curro}}, \bibinfo
  {author} {\bibfnamefont {J.~D.}\ \bibnamefont {Thompson}}, \bibinfo {author}
  {\bibfnamefont {J.~L.}\ \bibnamefont {Sarrao}}, \bibinfo {author}
  {\bibfnamefont {A.~B.}\ \bibnamefont {Vorontsov}}, \ and\ \bibinfo {author}
  {\bibfnamefont {M.~J.}\ \bibnamefont {Graf}},\ }\href {\doibase
  10.1103/PhysRevLett.98.036402} {\bibfield  {journal} {\bibinfo  {journal}
  {Phys. Rev. Lett.}\ }\textbf {\bibinfo {volume} {98}},\ \bibinfo {pages}
  {036402} (\bibinfo {year} {2007})}\BibitemShut {NoStop}%
\bibitem [{\citenamefont {Kenzelmann}\ \emph {et~al.}(2008)\citenamefont
  {Kenzelmann}, \citenamefont {Str{\"{a}}ssle}, \citenamefont {Niedermayer},
  \citenamefont {Sigrist}, \citenamefont {Padmanabhan}, \citenamefont
  {Zolliker}, \citenamefont {Bianchi}, \citenamefont {Movshovich},
  \citenamefont {Bauer}, \citenamefont {Sarrao},\ and\ \citenamefont
  {Thompson}}]{Kenzelmann2008}%
  \BibitemOpen
  \bibfield  {author} {\bibinfo {author} {\bibfnamefont {M.}~\bibnamefont
  {Kenzelmann}}, \bibinfo {author} {\bibfnamefont {T.}~\bibnamefont
  {Str{\"{a}}ssle}}, \bibinfo {author} {\bibfnamefont {C.}~\bibnamefont
  {Niedermayer}}, \bibinfo {author} {\bibfnamefont {M.}~\bibnamefont
  {Sigrist}}, \bibinfo {author} {\bibfnamefont {B.}~\bibnamefont
  {Padmanabhan}}, \bibinfo {author} {\bibfnamefont {M.}~\bibnamefont
  {Zolliker}}, \bibinfo {author} {\bibfnamefont {A.~D.}\ \bibnamefont
  {Bianchi}}, \bibinfo {author} {\bibfnamefont {R.}~\bibnamefont {Movshovich}},
  \bibinfo {author} {\bibfnamefont {E.~D.}\ \bibnamefont {Bauer}}, \bibinfo
  {author} {\bibfnamefont {J.~L.}\ \bibnamefont {Sarrao}}, \ and\ \bibinfo
  {author} {\bibfnamefont {J.~D.}\ \bibnamefont {Thompson}},\ }\href {\doibase
  10.1126/science.1161818} {\bibfield  {journal} {\bibinfo  {journal}
  {Science}\ }\textbf {\bibinfo {volume} {321}},\ \bibinfo {pages} {1652}
  (\bibinfo {year} {2008})}\BibitemShut {NoStop}%
\bibitem [{\citenamefont {Kenzelmann}\ \emph {et~al.}(2010)\citenamefont
  {Kenzelmann}, \citenamefont {Gerber}, \citenamefont {Egetenmeyer},
  \citenamefont {Gavilano}, \citenamefont {Str{\"{a}}ssle}, \citenamefont
  {Bianchi}, \citenamefont {Ressouche}, \citenamefont {Movshovich},
  \citenamefont {Bauer}, \citenamefont {Sarrao},\ and\ \citenamefont
  {Thompson}}]{Kenzelmann2010}%
  \BibitemOpen
  \bibfield  {author} {\bibinfo {author} {\bibfnamefont {M.}~\bibnamefont
  {Kenzelmann}}, \bibinfo {author} {\bibfnamefont {S.}~\bibnamefont {Gerber}},
  \bibinfo {author} {\bibfnamefont {N.}~\bibnamefont {Egetenmeyer}}, \bibinfo
  {author} {\bibfnamefont {J.~L.}\ \bibnamefont {Gavilano}}, \bibinfo {author}
  {\bibfnamefont {T.}~\bibnamefont {Str{\"{a}}ssle}}, \bibinfo {author}
  {\bibfnamefont {A.~D.}\ \bibnamefont {Bianchi}}, \bibinfo {author}
  {\bibfnamefont {E.}~\bibnamefont {Ressouche}}, \bibinfo {author}
  {\bibfnamefont {R.}~\bibnamefont {Movshovich}}, \bibinfo {author}
  {\bibfnamefont {E.~D.}\ \bibnamefont {Bauer}}, \bibinfo {author}
  {\bibfnamefont {J.~L.}\ \bibnamefont {Sarrao}}, \ and\ \bibinfo {author}
  {\bibfnamefont {J.~D.}\ \bibnamefont {Thompson}},\ }\href {\doibase
  10.1103/PhysRevLett.104.127001} {\bibfield  {journal} {\bibinfo  {journal}
  {Phys. Rev. Lett.}\ }\textbf {\bibinfo {volume} {104}},\ \bibinfo {pages}
  {127001} (\bibinfo {year} {2010})}\BibitemShut {NoStop}%
\bibitem [{\citenamefont {Agterberg}\ \emph {et~al.}(2009)\citenamefont
  {Agterberg}, \citenamefont {Sigrist},\ and\ \citenamefont
  {Tsunetsugu}}]{Agterberg}%
  \BibitemOpen
  \bibfield  {author} {\bibinfo {author} {\bibfnamefont {D.~F.}\ \bibnamefont
  {Agterberg}}, \bibinfo {author} {\bibfnamefont {M.}~\bibnamefont {Sigrist}},
  \ and\ \bibinfo {author} {\bibfnamefont {H.}~\bibnamefont {Tsunetsugu}},\
  }\href {\doibase 10.1103/PhysRevLett.102.207004} {\bibfield  {journal}
  {\bibinfo  {journal} {Phys. Rev. Lett.}\ }\textbf {\bibinfo {volume} {102}},\
  \bibinfo {pages} {207004} (\bibinfo {year} {2009})}\BibitemShut {NoStop}%
\bibitem [{\citenamefont {Aperis}\ \emph {et~al.}(2010)\citenamefont {Aperis},
  \citenamefont {Varelogiannis},\ and\ \citenamefont {Littlewood}}]{Aperis}%
  \BibitemOpen
  \bibfield  {author} {\bibinfo {author} {\bibfnamefont {A.}~\bibnamefont
  {Aperis}}, \bibinfo {author} {\bibfnamefont {G.}~\bibnamefont
  {Varelogiannis}}, \ and\ \bibinfo {author} {\bibfnamefont {P.~B.}\
  \bibnamefont {Littlewood}},\ }\href {\doibase 10.1103/PhysRevLett.104.216403}
  {\bibfield  {journal} {\bibinfo  {journal} {Phys. Rev. Lett.}\ }\textbf
  {\bibinfo {volume} {104}},\ \bibinfo {pages} {216403} (\bibinfo {year}
  {2010})}\BibitemShut {NoStop}%
\bibitem [{\citenamefont {Yanase}\ and\ \citenamefont
  {Sigrist}(2009)}]{Yanase2009}%
  \BibitemOpen
  \bibfield  {author} {\bibinfo {author} {\bibfnamefont {Y.}~\bibnamefont
  {Yanase}}\ and\ \bibinfo {author} {\bibfnamefont {M.}~\bibnamefont
  {Sigrist}},\ }\href {\doibase 10.1143/JPSJ.78.114715} {\bibfield  {journal}
  {\bibinfo  {journal} {J. Phys. Soc. Jpn.}\ }\textbf {\bibinfo {volume}
  {78}},\ \bibinfo {pages} {114715} (\bibinfo {year} {2009})}\BibitemShut
  {NoStop}%
\bibitem [{\citenamefont {Yanase}\ and\ \citenamefont
  {Sigrist}(2011)}]{Yanase2011}%
  \BibitemOpen
  \bibfield  {author} {\bibinfo {author} {\bibfnamefont {Y.}~\bibnamefont
  {Yanase}}\ and\ \bibinfo {author} {\bibfnamefont {M.}~\bibnamefont
  {Sigrist}},\ }\href {http://stacks.iop.org/0953-8984/23/i=9/a=094219}
  {\bibfield  {journal} {\bibinfo  {journal} {J. Phys. Condens. Matter}\
  }\textbf {\bibinfo {volume} {23}},\ \bibinfo {pages} {094219} (\bibinfo
  {year} {2011})}\BibitemShut {NoStop}%
\bibitem [{\citenamefont {Michal}\ and\ \citenamefont
  {Mineev}(2011)}]{Michal2011}%
  \BibitemOpen
  \bibfield  {author} {\bibinfo {author} {\bibfnamefont {V.~P.}\ \bibnamefont
  {Michal}}\ and\ \bibinfo {author} {\bibfnamefont {V.~P.}\ \bibnamefont
  {Mineev}},\ }\href {\doibase 10.1103/PhysRevB.84.052508} {\bibfield
  {journal} {\bibinfo  {journal} {Phys. Rev. B}\ }\textbf {\bibinfo {volume}
  {84}},\ \bibinfo {pages} {052508} (\bibinfo {year} {2011})}\BibitemShut
  {NoStop}%
\bibitem [{\citenamefont {Hatakeyama}\ and\ \citenamefont
  {Ikeda}(2011)}]{Hatakeyama2011}%
  \BibitemOpen
  \bibfield  {author} {\bibinfo {author} {\bibfnamefont {Y.}~\bibnamefont
  {Hatakeyama}}\ and\ \bibinfo {author} {\bibfnamefont {R.}~\bibnamefont
  {Ikeda}},\ }\href {\doibase 10.1103/PhysRevB.83.224518} {\bibfield  {journal}
  {\bibinfo  {journal} {Phys. Rev. B}\ }\textbf {\bibinfo {volume} {83}},\
  \bibinfo {pages} {224518} (\bibinfo {year} {2011})}\BibitemShut {NoStop}%
\bibitem [{\citenamefont {Martiny}\ \emph {et~al.}(2015)\citenamefont
  {Martiny}, \citenamefont {Gastiasoro}, \citenamefont {Vekhter},\ and\
  \citenamefont {Andersen}}]{Martiny}%
  \BibitemOpen
  \bibfield  {author} {\bibinfo {author} {\bibfnamefont {J.~H.~J.}\
  \bibnamefont {Martiny}}, \bibinfo {author} {\bibfnamefont {M.~N.}\
  \bibnamefont {Gastiasoro}}, \bibinfo {author} {\bibfnamefont
  {I.}~\bibnamefont {Vekhter}}, \ and\ \bibinfo {author} {\bibfnamefont
  {B.~M.}\ \bibnamefont {Andersen}},\ }\href {\doibase
  10.1103/PhysRevB.92.224510} {\bibfield  {journal} {\bibinfo  {journal} {Phys.
  Rev. B}\ }\textbf {\bibinfo {volume} {92}},\ \bibinfo {pages} {224510}
  (\bibinfo {year} {2015})}\BibitemShut {NoStop}%
\bibitem [{\citenamefont {Izawa}\ \emph {et~al.}(2001)\citenamefont {Izawa},
  \citenamefont {Yamaguchi}, \citenamefont {Matsuda}, \citenamefont {Shishido},
  \citenamefont {Settai},\ and\ \citenamefont {Onuki}}]{Izawa2001}%
  \BibitemOpen
  \bibfield  {author} {\bibinfo {author} {\bibfnamefont {K.}~\bibnamefont
  {Izawa}}, \bibinfo {author} {\bibfnamefont {H.}~\bibnamefont {Yamaguchi}},
  \bibinfo {author} {\bibfnamefont {Y.}~\bibnamefont {Matsuda}}, \bibinfo
  {author} {\bibfnamefont {H.}~\bibnamefont {Shishido}}, \bibinfo {author}
  {\bibfnamefont {R.}~\bibnamefont {Settai}}, \ and\ \bibinfo {author}
  {\bibfnamefont {Y.}~\bibnamefont {Onuki}},\ }\href {\doibase
  10.1103/PhysRevLett.87.057002} {\bibfield  {journal} {\bibinfo  {journal}
  {Phys. Rev. Lett.}\ }\textbf {\bibinfo {volume} {87}},\ \bibinfo {pages}
  {057002} (\bibinfo {year} {2001})}\BibitemShut {NoStop}%
\bibitem [{\citenamefont {Zhou}\ \emph {et~al.}(2013)\citenamefont {Zhou},
  \citenamefont {Misra}, \citenamefont {{da Silva Neto}}, \citenamefont
  {Aynajian}, \citenamefont {Baumbach}, \citenamefont {Thompson}, \citenamefont
  {Bauer},\ and\ \citenamefont {Yazdani}}]{Zhou2013}%
  \BibitemOpen
  \bibfield  {author} {\bibinfo {author} {\bibfnamefont {B.~B.}\ \bibnamefont
  {Zhou}}, \bibinfo {author} {\bibfnamefont {S.}~\bibnamefont {Misra}},
  \bibinfo {author} {\bibfnamefont {E.~H.}\ \bibnamefont {{da Silva Neto}}},
  \bibinfo {author} {\bibfnamefont {P.}~\bibnamefont {Aynajian}}, \bibinfo
  {author} {\bibfnamefont {R.~E.}\ \bibnamefont {Baumbach}}, \bibinfo {author}
  {\bibfnamefont {J.~D.}\ \bibnamefont {Thompson}}, \bibinfo {author}
  {\bibfnamefont {E.~D.}\ \bibnamefont {Bauer}}, \ and\ \bibinfo {author}
  {\bibfnamefont {A.}~\bibnamefont {Yazdani}},\ }\href
  {http://dx.doi.org/10.1038/nphys2672 http://10.1038/nphys2672
  http://www.nature.com/nphys/journal/v9/n8/abs/nphys2672.html#supplementary-information}
  {\bibfield  {journal} {\bibinfo  {journal} {Nat. Phys.}\ }\textbf {\bibinfo
  {volume} {9}},\ \bibinfo {pages} {474} (\bibinfo {year} {2013})}\BibitemShut
  {NoStop}%
\bibitem [{\citenamefont {Fradkin}\ \emph {et~al.}(2015)\citenamefont
  {Fradkin}, \citenamefont {Kivelson},\ and\ \citenamefont
  {Tranquada}}]{Fradkin2015}%
  \BibitemOpen
  \bibfield  {author} {\bibinfo {author} {\bibfnamefont {E.}~\bibnamefont
  {Fradkin}}, \bibinfo {author} {\bibfnamefont {S.~A.}\ \bibnamefont
  {Kivelson}}, \ and\ \bibinfo {author} {\bibfnamefont {J.~M.}\ \bibnamefont
  {Tranquada}},\ }\href {\doibase 10.1103/RevModPhys.87.457} {\bibfield
  {journal} {\bibinfo  {journal} {Rev. Mod. Phys.}\ }\textbf {\bibinfo {volume}
  {87}},\ \bibinfo {pages} {457} (\bibinfo {year} {2015})}\BibitemShut
  {NoStop}%
\bibitem [{\citenamefont {Hamidian}\ \emph {et~al.}(2016)\citenamefont
  {Hamidian}, \citenamefont {Edkins}, \citenamefont {Joo}, \citenamefont
  {Kostin}, \citenamefont {Eisaki}, \citenamefont {Uchida}, \citenamefont
  {Lawler}, \citenamefont {Kim}, \citenamefont {Mackenzie}, \citenamefont
  {Fujita}, \citenamefont {Lee},\ and\ \citenamefont {Davis}}]{Hamidian2016}%
  \BibitemOpen
  \bibfield  {author} {\bibinfo {author} {\bibfnamefont {M.~H.}\ \bibnamefont
  {Hamidian}}, \bibinfo {author} {\bibfnamefont {S.~D.}\ \bibnamefont
  {Edkins}}, \bibinfo {author} {\bibfnamefont {S.~H.}\ \bibnamefont {Joo}},
  \bibinfo {author} {\bibfnamefont {A.}~\bibnamefont {Kostin}}, \bibinfo
  {author} {\bibfnamefont {H.}~\bibnamefont {Eisaki}}, \bibinfo {author}
  {\bibfnamefont {S.}~\bibnamefont {Uchida}}, \bibinfo {author} {\bibfnamefont
  {M.~J.}\ \bibnamefont {Lawler}}, \bibinfo {author} {\bibfnamefont {E.-A.}\
  \bibnamefont {Kim}}, \bibinfo {author} {\bibfnamefont {A.~P.}\ \bibnamefont
  {Mackenzie}}, \bibinfo {author} {\bibfnamefont {K.}~\bibnamefont {Fujita}},
  \bibinfo {author} {\bibfnamefont {J.}~\bibnamefont {Lee}}, \ and\ \bibinfo
  {author} {\bibfnamefont {J.~C.~S.}\ \bibnamefont {Davis}},\ }\href
  {http://dx.doi.org/10.1038/nature17411 http://10.1038/nature17411} {\bibfield
   {journal} {\bibinfo  {journal} {Nature}\ }\textbf {\bibinfo {volume}
  {532}},\ \bibinfo {pages} {343} (\bibinfo {year} {2016})}\BibitemShut
  {NoStop}%
\bibitem [{\citenamefont {Stock}\ \emph {et~al.}(2008)\citenamefont {Stock},
  \citenamefont {Broholm}, \citenamefont {Hudis}, \citenamefont {Kang},\ and\
  \citenamefont {Petrovic}}]{Stock2008}%
  \BibitemOpen
  \bibfield  {author} {\bibinfo {author} {\bibfnamefont {C.}~\bibnamefont
  {Stock}}, \bibinfo {author} {\bibfnamefont {C.}~\bibnamefont {Broholm}},
  \bibinfo {author} {\bibfnamefont {J.}~\bibnamefont {Hudis}}, \bibinfo
  {author} {\bibfnamefont {H.~J.}\ \bibnamefont {Kang}}, \ and\ \bibinfo
  {author} {\bibfnamefont {C.}~\bibnamefont {Petrovic}},\ }\href {\doibase
  10.1103/PhysRevLett.100.087001} {\bibfield  {journal} {\bibinfo  {journal}
  {Phys. Rev. Lett.}\ }\textbf {\bibinfo {volume} {100}},\ \bibinfo {pages}
  {087001} (\bibinfo {year} {2008})}\BibitemShut {NoStop}%
\bibitem [{\citenamefont {Raymond}\ and\ \citenamefont
  {Lapertot}(2015)}]{Raymond2015}%
  \BibitemOpen
  \bibfield  {author} {\bibinfo {author} {\bibfnamefont {S.}~\bibnamefont
  {Raymond}}\ and\ \bibinfo {author} {\bibfnamefont {G.}~\bibnamefont
  {Lapertot}},\ }\href {\doibase 10.1103/PhysRevLett.115.037001} {\bibfield
  {journal} {\bibinfo  {journal} {Phys. Rev. Lett.}\ }\textbf {\bibinfo
  {volume} {115}},\ \bibinfo {pages} {037001} (\bibinfo {year}
  {2015})}\BibitemShut {NoStop}%
\bibitem [{\citenamefont {Gerber}\ \emph {et~al.}(2014)\citenamefont {Gerber},
  \citenamefont {Bartkowiak}, \citenamefont {Gavilano}, \citenamefont
  {Ressouche}, \citenamefont {Egetenmeyer}, \citenamefont {Niedermayer},
  \citenamefont {Bianchi}, \citenamefont {Movshovich}, \citenamefont {Bauer},
  \citenamefont {Thompson},\ and\ \citenamefont {Kenzelmann}}]{Gerber2014}%
  \BibitemOpen
  \bibfield  {author} {\bibinfo {author} {\bibfnamefont {S.}~\bibnamefont
  {Gerber}}, \bibinfo {author} {\bibfnamefont {M.}~\bibnamefont {Bartkowiak}},
  \bibinfo {author} {\bibfnamefont {J.~L.}\ \bibnamefont {Gavilano}}, \bibinfo
  {author} {\bibfnamefont {E.}~\bibnamefont {Ressouche}}, \bibinfo {author}
  {\bibfnamefont {N.}~\bibnamefont {Egetenmeyer}}, \bibinfo {author}
  {\bibfnamefont {C.}~\bibnamefont {Niedermayer}}, \bibinfo {author}
  {\bibfnamefont {A.~D.}\ \bibnamefont {Bianchi}}, \bibinfo {author}
  {\bibfnamefont {R.}~\bibnamefont {Movshovich}}, \bibinfo {author}
  {\bibfnamefont {E.~D.}\ \bibnamefont {Bauer}}, \bibinfo {author}
  {\bibfnamefont {J.~D.}\ \bibnamefont {Thompson}}, \ and\ \bibinfo {author}
  {\bibfnamefont {M.}~\bibnamefont {Kenzelmann}},\ }\href {\doibase
  10.1038/nphys2833} {\bibfield  {journal} {\bibinfo  {journal} {Nat. Phys.}\
  }\textbf {\bibinfo {volume} {10}},\ \bibinfo {pages} {126} (\bibinfo {year}
  {2014})}\BibitemShut {NoStop}%
\bibitem [{\citenamefont {Mineev}(2015)}]{Mineev2015}%
  \BibitemOpen
  \bibfield  {author} {\bibinfo {author} {\bibfnamefont {V.}~\bibnamefont
  {Mineev}},\ }\href@noop {} {\bibfield  {journal} {\bibinfo  {journal}
  {arXiv:1509.04915}\ } (\bibinfo {year} {2015})}\BibitemShut {NoStop}%
\bibitem [{\citenamefont {Hatakeyama}\ and\ \citenamefont
  {Ikeda}(2015)}]{Hatakeyama2015}%
  \BibitemOpen
  \bibfield  {author} {\bibinfo {author} {\bibfnamefont {Y.}~\bibnamefont
  {Hatakeyama}}\ and\ \bibinfo {author} {\bibfnamefont {R.}~\bibnamefont
  {Ikeda}},\ }\href {\doibase 10.1103/PhysRevB.91.094504} {\bibfield  {journal}
  {\bibinfo  {journal} {Phys. Rev. B}\ }\textbf {\bibinfo {volume} {91}},\
  \bibinfo {pages} {094504} (\bibinfo {year} {2015})}\BibitemShut {NoStop}%
\bibitem [{\citenamefont {Matsuda}\ \emph {et~al.}(2006)\citenamefont
  {Matsuda}, \citenamefont {Izawa},\ and\ \citenamefont
  {Vekhter}}]{Matsuda2006}%
  \BibitemOpen
  \bibfield  {author} {\bibinfo {author} {\bibfnamefont {Y.}~\bibnamefont
  {Matsuda}}, \bibinfo {author} {\bibfnamefont {K.}~\bibnamefont {Izawa}}, \
  and\ \bibinfo {author} {\bibfnamefont {I.}~\bibnamefont {Vekhter}},\ }\href
  {http://stacks.iop.org/0953-8984/18/i=44/a=R01} {\bibfield  {journal}
  {\bibinfo  {journal} {J. Phys. Condens. Matter}\ }\textbf {\bibinfo {volume}
  {18}},\ \bibinfo {pages} {R705} (\bibinfo {year} {2006})}\BibitemShut
  {NoStop}%
\bibitem [{\citenamefont {Shakeripour}\ \emph {et~al.}(2009)\citenamefont
  {Shakeripour}, \citenamefont {Petrovic},\ and\ \citenamefont
  {Taillefer}}]{Shakeripour}%
  \BibitemOpen
  \bibfield  {author} {\bibinfo {author} {\bibfnamefont {H.}~\bibnamefont
  {Shakeripour}}, \bibinfo {author} {\bibfnamefont {C.}~\bibnamefont
  {Petrovic}}, \ and\ \bibinfo {author} {\bibfnamefont {L.}~\bibnamefont
  {Taillefer}},\ }\href {http://stacks.iop.org/1367-2630/11/i=5/a=055065}
  {\bibfield  {journal} {\bibinfo  {journal} {New J. Phys.}\ }\textbf {\bibinfo
  {volume} {11}},\ \bibinfo {pages} {055065} (\bibinfo {year}
  {2009})}\BibitemShut {NoStop}%
\bibitem [{\citenamefont {Yeoh}\ \emph {et~al.}(2010)\citenamefont {Yeoh},
  \citenamefont {Srinivasan}, \citenamefont {Martin}, \citenamefont {Klochan},
  \citenamefont {Micolich},\ and\ \citenamefont {Hamilton}}]{Yeoh}%
  \BibitemOpen
  \bibfield  {author} {\bibinfo {author} {\bibfnamefont {L.~A.}\ \bibnamefont
  {Yeoh}}, \bibinfo {author} {\bibfnamefont {A.}~\bibnamefont {Srinivasan}},
  \bibinfo {author} {\bibfnamefont {T.~P.}\ \bibnamefont {Martin}}, \bibinfo
  {author} {\bibfnamefont {O.}~\bibnamefont {Klochan}}, \bibinfo {author}
  {\bibfnamefont {A.~P.}\ \bibnamefont {Micolich}}, \ and\ \bibinfo {author}
  {\bibfnamefont {A.~R.}\ \bibnamefont {Hamilton}},\ }\href
  {http://scitation.aip.org/content/aip/journal/rsi/81/11/10.1063/1.3502645}
  {\bibfield  {journal} {\bibinfo  {journal} {Rev. Sci. Instrum.}\ }\textbf
  {\bibinfo {volume} {81}},\ \bibinfo {pages} {113905} (\bibinfo {year}
  {2010})}\BibitemShut {NoStop}%
\bibitem [{\citenamefont {Feldman}\ \emph {et~al.}(2016)\citenamefont
  {Feldman}, \citenamefont {Gyenis}, \citenamefont {Randeria}, \citenamefont
  {Peterson}, \citenamefont {Aynajian}, \citenamefont {Bauer},\ and\
  \citenamefont {Yazdani}}]{Feldman2016}%
  \BibitemOpen
  \bibfield  {author} {\bibinfo {author} {\bibfnamefont {B.~E.}\ \bibnamefont
  {Feldman}}, \bibinfo {author} {\bibfnamefont {A.}~\bibnamefont {Gyenis}},
  \bibinfo {author} {\bibfnamefont {M.~T.}\ \bibnamefont {Randeria}}, \bibinfo
  {author} {\bibfnamefont {G.~A.}\ \bibnamefont {Peterson}}, \bibinfo {author}
  {\bibfnamefont {P.}~\bibnamefont {Aynajian}}, \bibinfo {author}
  {\bibfnamefont {E.~D.}\ \bibnamefont {Bauer}}, \ and\ \bibinfo {author}
  {\bibfnamefont {A.}~\bibnamefont {Yazdani}},\ }\href
  {http://meetings.aps.org/link/BAPS.2016.MAR.E27.10} {\bibfield  {journal}
  {\bibinfo  {journal} {Bull. Am. Phys. Soc.}\ } (\bibinfo {year}
  {2016})}\BibitemShut {NoStop}%
\bibitem [{\citenamefont {Das}\ \emph {et~al.}(2012)\citenamefont {Das},
  \citenamefont {White}, \citenamefont {Holmes}, \citenamefont {Gerber},
  \citenamefont {Forgan}, \citenamefont {Bianchi}, \citenamefont {Kenzelmann},
  \citenamefont {Zolliker}, \citenamefont {Gavilano}, \citenamefont {Bauer},
  \citenamefont {Sarrao}, \citenamefont {Petrovic},\ and\ \citenamefont
  {Eskildsen}}]{Das2012}%
  \BibitemOpen
  \bibfield  {author} {\bibinfo {author} {\bibfnamefont {P.}~\bibnamefont
  {Das}}, \bibinfo {author} {\bibfnamefont {J.~S.}\ \bibnamefont {White}},
  \bibinfo {author} {\bibfnamefont {A.~T.}\ \bibnamefont {Holmes}}, \bibinfo
  {author} {\bibfnamefont {S.}~\bibnamefont {Gerber}}, \bibinfo {author}
  {\bibfnamefont {E.~M.}\ \bibnamefont {Forgan}}, \bibinfo {author}
  {\bibfnamefont {A.~D.}\ \bibnamefont {Bianchi}}, \bibinfo {author}
  {\bibfnamefont {M.}~\bibnamefont {Kenzelmann}}, \bibinfo {author}
  {\bibfnamefont {M.}~\bibnamefont {Zolliker}}, \bibinfo {author}
  {\bibfnamefont {J.~L.}\ \bibnamefont {Gavilano}}, \bibinfo {author}
  {\bibfnamefont {E.~D.}\ \bibnamefont {Bauer}}, \bibinfo {author}
  {\bibfnamefont {J.~L.}\ \bibnamefont {Sarrao}}, \bibinfo {author}
  {\bibfnamefont {C.}~\bibnamefont {Petrovic}}, \ and\ \bibinfo {author}
  {\bibfnamefont {M.~R.}\ \bibnamefont {Eskildsen}},\ }\href {\doibase
  10.1103/PhysRevLett.108.087002} {\bibfield  {journal} {\bibinfo  {journal}
  {Phys. Rev. Lett.}\ }\textbf {\bibinfo {volume} {108}},\ \bibinfo {pages}
  {087002} (\bibinfo {year} {2012})}\BibitemShut {NoStop}%
\bibitem [{\citenamefont {Paglione}\ \emph {et~al.}(2016)\citenamefont
  {Paglione}, \citenamefont {Tanatar}, \citenamefont {Reid}, \citenamefont
  {Shakeripour}, \citenamefont {Petrovic},\ and\ \citenamefont
  {Taillefer}}]{Paglione2016}%
  \BibitemOpen
  \bibfield  {author} {\bibinfo {author} {\bibfnamefont {J.}~\bibnamefont
  {Paglione}}, \bibinfo {author} {\bibfnamefont {M.~A.}\ \bibnamefont
  {Tanatar}}, \bibinfo {author} {\bibfnamefont {J.-P.}\ \bibnamefont {Reid}},
  \bibinfo {author} {\bibfnamefont {H.}~\bibnamefont {Shakeripour}}, \bibinfo
  {author} {\bibfnamefont {C.}~\bibnamefont {Petrovic}}, \ and\ \bibinfo
  {author} {\bibfnamefont {L.}~\bibnamefont {Taillefer}},\ }\href {\doibase
  10.1103/PhysRevLett.117.016601} {\bibfield  {journal} {\bibinfo  {journal}
  {Phys. Rev. Lett.}\ }\textbf {\bibinfo {volume} {117}},\ \bibinfo {pages}
  {016601} (\bibinfo {year} {2016})}\BibitemShut {NoStop}%
\bibitem [{\citenamefont {Ronning}\ \emph {et~al.}(2005)\citenamefont
  {Ronning}, \citenamefont {Capan}, \citenamefont {Bianchi}, \citenamefont
  {Movshovich}, \citenamefont {Lacerda}, \citenamefont {Hundley}, \citenamefont
  {Thompson}, \citenamefont {Pagliuso},\ and\ \citenamefont
  {Sarrao}}]{Ronning2005}%
  \BibitemOpen
  \bibfield  {author} {\bibinfo {author} {\bibfnamefont {F.}~\bibnamefont
  {Ronning}}, \bibinfo {author} {\bibfnamefont {C.}~\bibnamefont {Capan}},
  \bibinfo {author} {\bibfnamefont {A.}~\bibnamefont {Bianchi}}, \bibinfo
  {author} {\bibfnamefont {R.}~\bibnamefont {Movshovich}}, \bibinfo {author}
  {\bibfnamefont {A.}~\bibnamefont {Lacerda}}, \bibinfo {author} {\bibfnamefont
  {M.~F.}\ \bibnamefont {Hundley}}, \bibinfo {author} {\bibfnamefont {J.~D.}\
  \bibnamefont {Thompson}}, \bibinfo {author} {\bibfnamefont {P.~G.}\
  \bibnamefont {Pagliuso}}, \ and\ \bibinfo {author} {\bibfnamefont {J.~L.}\
  \bibnamefont {Sarrao}},\ }\href {\doibase 10.1103/PhysRevB.71.104528}
  {\bibfield  {journal} {\bibinfo  {journal} {Phys. Rev. B}\ }\textbf {\bibinfo
  {volume} {71}},\ \bibinfo {pages} {104528} (\bibinfo {year}
  {2005})}\BibitemShut {NoStop}%
\bibitem [{\citenamefont {Koutroulakis}\ \emph {et~al.}(2010)\citenamefont
  {Koutroulakis}, \citenamefont {Stewart}, \citenamefont {Mitrovi{\'{c}}},
  \citenamefont {Horvati{\'{c}}}, \citenamefont {Berthier}, \citenamefont
  {Lapertot},\ and\ \citenamefont {Flouquet}}]{Koutroulakis2010}%
  \BibitemOpen
  \bibfield  {author} {\bibinfo {author} {\bibfnamefont {G.}~\bibnamefont
  {Koutroulakis}}, \bibinfo {author} {\bibfnamefont {M.~D.}\ \bibnamefont
  {Stewart}}, \bibinfo {author} {\bibfnamefont {V.~F.}\ \bibnamefont
  {Mitrovi{\'{c}}}}, \bibinfo {author} {\bibfnamefont {M.}~\bibnamefont
  {Horvati{\'{c}}}}, \bibinfo {author} {\bibfnamefont {C.}~\bibnamefont
  {Berthier}}, \bibinfo {author} {\bibfnamefont {G.}~\bibnamefont {Lapertot}},
  \ and\ \bibinfo {author} {\bibfnamefont {J.}~\bibnamefont {Flouquet}},\
  }\href {\doibase 10.1103/PhysRevLett.104.087001} {\bibfield  {journal}
  {\bibinfo  {journal} {Phys. Rev. Lett.}\ }\textbf {\bibinfo {volume} {104}},\
  \bibinfo {pages} {087001} (\bibinfo {year} {2010})}\BibitemShut {NoStop}%
\bibitem [{\citenamefont {Vorontsov}\ and\ \citenamefont
  {Vekhter}(2007)}]{Vorontsov2007}%
  \BibitemOpen
  \bibfield  {author} {\bibinfo {author} {\bibfnamefont {A.~B.}\ \bibnamefont
  {Vorontsov}}\ and\ \bibinfo {author} {\bibfnamefont {I.}~\bibnamefont
  {Vekhter}},\ }\href {\doibase 10.1103/PhysRevB.75.224502} {\bibfield
  {journal} {\bibinfo  {journal} {Phys. Rev. B}\ }\textbf {\bibinfo {volume}
  {75}},\ \bibinfo {pages} {224502} (\bibinfo {year} {2007})}\BibitemShut
  {NoStop}%
\bibitem [{\citenamefont {Maki}(1966)}]{Maki}%
  \BibitemOpen
  \bibfield  {author} {\bibinfo {author} {\bibfnamefont {K.}~\bibnamefont
  {Maki}},\ }\href {\doibase 10.1103/PhysRev.148.362} {\bibfield  {journal}
  {\bibinfo  {journal} {Phys. Rev.}\ }\textbf {\bibinfo {volume} {148}},\
  \bibinfo {pages} {362} (\bibinfo {year} {1966})}\BibitemShut {NoStop}%
\bibitem [{\citenamefont {Movshovich}\ \emph {et~al.}(2001)\citenamefont
  {Movshovich}, \citenamefont {Jaime}, \citenamefont {Thompson}, \citenamefont
  {Petrovic}, \citenamefont {Fisk}, \citenamefont {Pagliuso},\ and\
  \citenamefont {Sarrao}}]{Movshovich2001}%
  \BibitemOpen
  \bibfield  {author} {\bibinfo {author} {\bibfnamefont {R.}~\bibnamefont
  {Movshovich}}, \bibinfo {author} {\bibfnamefont {M.}~\bibnamefont {Jaime}},
  \bibinfo {author} {\bibfnamefont {J.~D.}\ \bibnamefont {Thompson}}, \bibinfo
  {author} {\bibfnamefont {C.}~\bibnamefont {Petrovic}}, \bibinfo {author}
  {\bibfnamefont {Z.}~\bibnamefont {Fisk}}, \bibinfo {author} {\bibfnamefont
  {P.~G.}\ \bibnamefont {Pagliuso}}, \ and\ \bibinfo {author} {\bibfnamefont
  {J.~L.}\ \bibnamefont {Sarrao}},\ }\href {\doibase
  10.1103/PhysRevLett.86.5152} {\bibfield  {journal} {\bibinfo  {journal}
  {Phys. Rev. Lett.}\ }\textbf {\bibinfo {volume} {86}},\ \bibinfo {pages}
  {5152} (\bibinfo {year} {2001})}\BibitemShut {NoStop}%
\bibitem [{\citenamefont {Kumagai}\ \emph {et~al.}(2011)\citenamefont
  {Kumagai}, \citenamefont {Shishido}, \citenamefont {Shibauchi},\ and\
  \citenamefont {Matsuda}}]{kumagai2011}%
  \BibitemOpen
  \bibfield  {author} {\bibinfo {author} {\bibfnamefont {K.}~\bibnamefont
  {Kumagai}}, \bibinfo {author} {\bibfnamefont {H.}~\bibnamefont {Shishido}},
  \bibinfo {author} {\bibfnamefont {T.}~\bibnamefont {Shibauchi}}, \ and\
  \bibinfo {author} {\bibfnamefont {Y.}~\bibnamefont {Matsuda}},\ }\href
  {\doibase 10.1103/PhysRevLett.106.137004} {\bibfield  {journal} {\bibinfo
  {journal} {Phys. Rev. Lett.}\ }\textbf {\bibinfo {volume} {106}},\ \bibinfo
  {pages} {137004} (\bibinfo {year} {2011})}\BibitemShut {NoStop}%
\bibitem [{\citenamefont {Tokiwa}\ \emph {et~al.}(2010)\citenamefont {Tokiwa},
  \citenamefont {Movshovich}, \citenamefont {Ronning}, \citenamefont {Bauer},
  \citenamefont {Bianchi}, \citenamefont {Fisk},\ and\ \citenamefont
  {Thompson}}]{Tokiwa2010}%
  \BibitemOpen
  \bibfield  {author} {\bibinfo {author} {\bibfnamefont {Y.}~\bibnamefont
  {Tokiwa}}, \bibinfo {author} {\bibfnamefont {R.}~\bibnamefont {Movshovich}},
  \bibinfo {author} {\bibfnamefont {F.}~\bibnamefont {Ronning}}, \bibinfo
  {author} {\bibfnamefont {E.~D.}\ \bibnamefont {Bauer}}, \bibinfo {author}
  {\bibfnamefont {A.~D.}\ \bibnamefont {Bianchi}}, \bibinfo {author}
  {\bibfnamefont {Z.}~\bibnamefont {Fisk}}, \ and\ \bibinfo {author}
  {\bibfnamefont {J.~D.}\ \bibnamefont {Thompson}},\ }\href {\doibase
  10.1103/PhysRevB.82.220502} {\bibfield  {journal} {\bibinfo  {journal} {Phys.
  Rev. B}\ }\textbf {\bibinfo {volume} {82}},\ \bibinfo {pages} {220502}
  (\bibinfo {year} {2010})}\BibitemShut {NoStop}%
\bibitem [{\citenamefont {Ikeda}(2010)}]{Ikeda2010}%
  \BibitemOpen
  \bibfield  {author} {\bibinfo {author} {\bibfnamefont {R.}~\bibnamefont
  {Ikeda}},\ }\href {\doibase 10.1103/PhysRevB.81.060510} {\bibfield  {journal}
  {\bibinfo  {journal} {Phys. Rev. B}\ }\textbf {\bibinfo {volume} {81}},\
  \bibinfo {pages} {060510} (\bibinfo {year} {2010})}\BibitemShut {NoStop}%
\bibitem [{\citenamefont {Vorontsov}\ \emph {et~al.}(2005)\citenamefont
  {Vorontsov}, \citenamefont {Sauls},\ and\ \citenamefont
  {Graf}}]{Vorontsov2005}%
  \BibitemOpen
  \bibfield  {author} {\bibinfo {author} {\bibfnamefont {A.~B.}\ \bibnamefont
  {Vorontsov}}, \bibinfo {author} {\bibfnamefont {J.~A.}\ \bibnamefont
  {Sauls}}, \ and\ \bibinfo {author} {\bibfnamefont {M.~J.}\ \bibnamefont
  {Graf}},\ }\href {\doibase 10.1103/PhysRevB.72.184501} {\bibfield  {journal}
  {\bibinfo  {journal} {Phys. Rev. B}\ }\textbf {\bibinfo {volume} {72}},\
  \bibinfo {pages} {184501} (\bibinfo {year} {2005})}\BibitemShut {NoStop}%
\bibitem [{\citenamefont {Bulaevskii}\ \emph {et~al.}(2016)\citenamefont
  {Bulaevskii}, \citenamefont {Eneias},\ and\ \citenamefont
  {Ferraz}}]{Bulaevskii}%
  \BibitemOpen
  \bibfield  {author} {\bibinfo {author} {\bibfnamefont {L.}~\bibnamefont
  {Bulaevskii}}, \bibinfo {author} {\bibfnamefont {R.}~\bibnamefont {Eneias}},
  \ and\ \bibinfo {author} {\bibfnamefont {A.}~\bibnamefont {Ferraz}},\ }\href
  {\doibase 10.1103/PhysRevB.93.014501} {\bibfield  {journal} {\bibinfo
  {journal} {Phys. Rev. B}\ }\textbf {\bibinfo {volume} {93}},\ \bibinfo
  {pages} {014501} (\bibinfo {year} {2016})}\BibitemShut {NoStop}%
\bibitem [{\citenamefont {Lee}\ \emph {et~al.}(1974)\citenamefont {Lee},
  \citenamefont {Rice},\ and\ \citenamefont {Anderson}}]{Lee}%
  \BibitemOpen
  \bibfield  {author} {\bibinfo {author} {\bibfnamefont {P.~A.}\ \bibnamefont
  {Lee}}, \bibinfo {author} {\bibfnamefont {T.~M.}\ \bibnamefont {Rice}}, \
  and\ \bibinfo {author} {\bibfnamefont {P.~W.}\ \bibnamefont {Anderson}},\
  }\href {\doibase http://dx.doi.org/10.1016/0038-1098(74)90868-0} {\bibfield
  {journal} {\bibinfo  {journal} {Solid State Commun.}\ }\textbf {\bibinfo
  {volume} {14}},\ \bibinfo {pages} {703} (\bibinfo {year} {1974})}\BibitemShut
  {NoStop}%
\end{thebibliography}%

\end{document}